\documentclass[aps,preprintnumbers,floatfix,nofootinbib,preprint,superscriptaddress]{revtex4}

\pdfoutput=1

\usepackage{mathrsfs}
\usepackage{amsmath, amssymb}
\usepackage{color}
\usepackage[pdftex]{graphicx}
\usepackage{amssymb}
\usepackage{verbatim}
\usepackage{hyperref}
\usepackage{multirow}
\usepackage[normalem]{ulem}

\newcommand{\be}{\begin{equation}}
\newcommand{\ee}{\end{equation}}
\newcommand{\bea}{\begin{eqnarray}}
\newcommand{\eea}{\end{eqnarray}}

\newcommand{\df}{\dfrac}

\newcommand{\ie}{i.e.}

\newcommand{\coff}[1]{c_{#1}}
\renewcommand{\c}[1]{\coff{#1}}
\newcommand{\cL}[1]{\coff{L #1}}
\newcommand{\cQ}[1]{\coff{Q #1}}
\newcommand{\cu}{\coff{uR}}
\newcommand{\cd}{\coff{dR}}
\newcommand{\cc}{\coff{cR}}
\newcommand{\cs}{\coff{sR}}
\newcommand{\ct}{\coff{tR}}
\newcommand{\cb}{\coff{bR}}
\newcommand{\ctau}{\coff{\tau R}}
\newcommand{\cmu}{\coff{\mu R}}
\newcommand{\ce}{\coff{eR}}
\newcommand{\cf}{\coff{fR}}

\newcommand{\micrO}{{\tt micrOmegas}}

\begin{document}

\preprint{\footnotesize FTUAM-15-27, IFT-UAM/CSIC-15-095, ULB-TH/15-17, FERMILAB-PUB-15-374-T}

\title{Global constraints on vector-like WIMP effective interactions}
\author{Mattias Blennow}
\email{emb@kth.se}
\affiliation{\footnotesize Department of Theoretical Physics, School of Engineering Sciences, KTH Royal Institute of Technology, 
  Albanova University Center, 106 91 Stockholm, Sweden}

\author{Pilar Coloma}
\email{pcoloma@fnal.gov}
\affiliation{\footnotesize Center for Neutrino Physics, Physics Department, Virginia Tech, \\
850 West Campus Dr, Blacksburg, VA 24061, USA  }
\affiliation{\footnotesize Theoretical Physics Department, Fermi National Accelerator Laboratory, \\
P.O. Box 500, Batavia, IL 60510, USA  }

\author{Enrique Fern\'andez-Mart\'inez}
\email{enrique.fernandez-martinez@uam.es}
\affiliation{\footnotesize Departamento de F\'isica Te\'orica, Universidad Aut\'onoma de Madrid, Cantoblanco E-28049 Madrid, Spain}
\affiliation{\footnotesize Instituto de F\'isica Te\'orica UAM/CSIC,
 Calle Nicol\'as Cabrera 13-15, Cantoblanco E-28049 Madrid, Spain}

\author{Pedro A. N. Machado}
\email{pedro.machado@uam.es}
\affiliation{\footnotesize Departamento de F\'isica Te\'orica, Universidad Aut\'onoma de Madrid, Cantoblanco E-28049 Madrid, Spain}
\affiliation{\footnotesize Instituto de F\'isica Te\'orica UAM/CSIC,
 Calle Nicol\'as Cabrera 13-15, Cantoblanco E-28049 Madrid, Spain}

\author{Bryan Zald\'\i var}
\email{bryan.zaldivar@ulb.ac.be}
\affiliation{\footnotesize Service de Physique Th\'eorique, Universit\'e Libre de Bruxelles, Boulevard du Triomphe, CP225, 1050 Brussels, Belgium.}

\begin{abstract}
In this work we combine information from relic abundance, direct
detection, cosmic microwave background, positron fraction, gamma rays,
and colliders to explore the existing constraints on couplings between
Dark Matter and Standard Model constituents when no underlying model
or correlation is assumed.  For definiteness, we include independent
vector-like effective interactions for each Standard Model fermion.
Our results show that low Dark Matter masses below 20~GeV are disfavoured
at the $3 \sigma$ level with respect to higher masses, due to the tension between the relic abundance
requirement and upper constraints on the Dark Matter
couplings. Furthermore, large couplings are typically only allowed in
combinations which avoid effective couplings to the nuclei used in
direct detection experiments.
\end{abstract}

\maketitle

\section{Introduction}

The search for Dark Matter (DM) in the form of thermal relics represents one of the most active lines of research in astro-particle and particle physics. Indeed, there is an overwhelming number of dedicated experimental searches for DM, most of them concentrating on the so-called Weakly Interacting Massive Particles (WIMPs) paradigm \cite{Bergstrom:2000pn}. These are classified into three different categories: (1) indirect detection searches, where the DM would annihilate or decay into SM particles which can be detected, (2) direct detection searches, where the DM would scatter against the protons and neutrons in a detector, producing an observable recoil, and (3) collider searches, where the DM would be produced in high-energy collisions, thus leading to events with missing momentum.\footnote{A complementary search strategy is the search for DM self-interactions which would impact structure formation as well as stellar evolution in particular scenarios (see e.g. \cite{Tulin:2013teo,Hochberg:2014dra,Vincent:2014jia,Vincent:2015gqa}). Since the focus of this work is to probe the interactions between DM and visible particles these constraints will not be considered here.}

In light of all these searches it is essential to look at the global picture of where we stand concerning the WIMP scenario. This is certainly a rather complicated task, given the enormous zoo of models available in the literature with an associated plethora of free parameters. Even for a single model, brute-force scans of the corresponding parameter space represent a significant computational challenge. Consequently, the usual practice is to either rely on simplified realisations of those models or to fix some of the free parameters, thus scanning only hyper-surfaces in the model parameter space.
\footnote{The literature in this respect is quite vast. Some examples relevant to our work are \cite{Beltran:2008xg,Beltran:2010ww,Fox:2011fx,Fox:2011pm,Mambrini:2011pw, Cheung:2012gi}, in the context of effective field theory (see below).} 

On the other hand, Monte Carlo methods constitute an efficient alternative to scan over a multi-dimensional parameter space. However, for WIMPs, most of the Monte Carlo scans are performed in the context of supersymmetric models, where the neutralino is usually selected as the preferred DM candidate (e.g., Refs.~\cite{Cabrera:2012vu,Roszkowski:2012uf,Strege:2014ija}, however see \cite{Cheung:2012xb} as an example of a non-SUSY search).

The main goal of this work is to explore the present status of our
knowledge of DM couplings to the different SM constituents in as much
generality as possible. In order to gain in model independence, it is
thus interesting to be more agnostic about the DM interactions with
the SM constituents and parametrize them through an effective field
theory (EFT) approach. In particular, this approach can reveal how
constrained the DM interactions with the different SM fields are in
the light of currently available data, when no further assumptions are
made. Since allowing all type of Lorentz structures for these
effective operators would provide too much freedom with hundreds of
operators that cannot be bounded, we rather choose to exemplify the
different interactions between a Dirac fermion DM and the SM by
operators of the form
\begin{equation}
c_{i,P} \left(\bar{\chi} \gamma^\mu \chi \right) \left(\bar{f_i} \gamma_\mu P f_i \right)~, 
\label{model0}
\end{equation}
with independent coefficients $c_{i,P}$ for all SM particle species
$f_i$ with chiralities $P \equiv P_L, P_R$. This type of interaction
is well motivated by scenarios where the DM communicates with the SM
via a vector portal.  Moreover, scalar or tensor interactions
typically require Higgs insertions, leading to higher dimensional and
hence more suppressed operators. Additional dimension six operators with a vector-like structure involving the Higgs, such as $i \bar \chi \gamma^\mu \chi H^\dagger \overleftrightarrow{D}_{\mu} H$, could be included as well. For DM heavier than the Higgs, this operator would provide an additional annihilation channel relevant for indirect searches and relic abundance constraints, via $\bar\chi \chi \rightarrow hh$. However, as we will see, above 100~GeV the global upper bounds for any individual coupling correspond to the ones obtained from relic abundance alone (with no constraint for the top quark coupling, since that channel is not kinematically allowed). Therefore, we expect a similar behaviour for the Higgs, that is, no constraints for DM masses below the Higgs mass and recovering the relic abundance constraint for higher DM masses. Even with this restriction, if
independent couplings for all SM constituents are allowed, 15
independent parameters remain to be constrained. Thus, given the large
dimensionality of the parameter space, we make use of a nested
sampling Monte Carlo algorithm to scan it efficiently. 

When constraining the DM EFT parameter space, we consider bounds from
all types of experiments where a WIMP signal is being actively sought
for, \ie, direct detection (namely, LUX \cite{Akerib:2013tjd} and
EDELWEISS \cite{Ahmed:2011gh}), indirect detection (AMS
\cite{Aguilar:2013qda} positron fraction data and Fermi-LAT data for
dwarf galaxies \cite{Ackermann:2015zua}), cosmic microwave background
(CMB) and relic density constraints from Planck \cite{Ade:2015xua},
and monojet and monophoton searches in colliders (from LHC
\cite{Aad:2015zva} and LEP data \cite{Fox:2011fx}).\footnote{Contrary to the cases of direct and indirect detection
  strategies, for collider searches the use of the EFT framework
  -mainly at the LHC- may not be optimal. We take this issue into
  account when recasting the limits coming from LHC. See
  Sec.\ref{sec:constraints}. Also, note that other searches
  different from monojets can be interesting in the context of
  specific models (see e.g. Ref.~\cite{CMS:2014gxa}.}  

Additional information on some annihilation channels such as
$\tau\tau$, $bb$, or $\nu\nu$ come from the non-observation of
indirect neutrino signals from the Sun in
SuperKamiokande~\cite{Choi:2015ara} and
IceCube~\cite{Aartsen:2016exj}. However, the bound is only relevant if
apart from a large DM coupling to the final state particle, a sizeable
coupling to first family quarks is also present. Since we are assuming
all couplings to be independent, this bound will always be satisfied in the fit by choosing a set of parameters with one of the relevant couplings being very small. Thus, the inclusion of these experiments
would not change our results.

We derive bounds on the coefficients accompanying each effective
operator as a function of the DM mass, as well as bounds on the DM
mass itself. These constitute the most general constraints which can
be derived assuming that DM interacts with the SM as in
Eq.~\eqref{model0}.

The outline of the work is as follows. In Sec.~\ref{sec:setups} we describe the set of effective operators that will be jointly analyzed, introducing the parameters to be constrained. Sec.~\ref{sec:constraints} lists the set of experimental constraints considered, and how these have been implemented. Some details regarding the numerical tools employed in the fit are explained in Sec.~\ref{sec:scan}. Finally, Sec.~\ref{sec:results} summarizes our results and we present our conclusions in Sec.~\ref{sec:concl}. 

\section{The Effective Field Theory Framework}
\label{sec:setups}

In order to explore how much freedom the present global data on DM allows for its couplings to the SM, we will define a series of working models characterized by a set of independent effective operators, described by the following effective Lagrangian
\begin{equation}
\mathcal L_{\rm eff} = (\overline \chi \gamma^\mu \chi) j^{\rm eff}_\mu.
\label{model1}
\end{equation}
where the effective current $j_\mu^{\rm eff}$ is given by 
\begin{equation}
j_\mu^{\rm eff} = \sum_{\ell = e, \mu, \tau} \cL\ell  \begin{pmatrix}
\overline{\nu_{\ell,L}} & \overline{\ell_L}
\end{pmatrix} \gamma_\mu \begin{pmatrix}
\nu_{\ell,L} \\ \ell_L
\end{pmatrix} +
\sum_{i=1}^3 \cQ i \begin{pmatrix}
\overline{u_{i,L}} & \overline{d_{i,L}}
\end{pmatrix} \gamma_\mu \begin{pmatrix}
u_{i,L} \\ d_{i,L}
\end{pmatrix}+
\sum_{f} \cf \overline{f_R} \gamma_\mu f_R,
\end{equation}
the index $i$ denotes the quark generations such that $u_1 = u$, $u_2 = c$, $u_3 = t$, $d_1 = d$, $d_2 = s$, and $d_3 = b$, while the sum over $f$ runs over all right-handed (RH) SM fermions. The coefficients $\c X$ are the couplings of the operators in the effective Lagrangian. As the effective operators are of dimension six, these coefficients will have a mass dimension of minus two. In expr.(\ref{model1}), $\chi$ represents a Dirac fermion DM.

This set of operators provide enough freedom so as to parameterize DM interactions with different strength to the various SM particles while, at the same time, keeping the set of operators at a viable level. Indeed, allowing extra operators would not only imply a more challenging numerical analysis, but would also be rather uninformative since mostly any value could be fitted and particular UV completions tend to have a much more limited set of free parameters. Furthermore, for simplicity and due to the generally stronger constraints, we will not consider flavour-changing operators between the SM fermion bilinears \cite{Kamenik:2011vy}. Despite these restrictions, 15 different operators fall into this category, parametrizing the DM interactions with the 3 quark and lepton doublets as well as the 3 singlets of up-type and down-type quarks and charged leptons. Starting from this general setup, we will also define more restrictive working models that can exemplify other interesting scenarios such as leptophilic, leptophobic or flavour-blind setups, characterized by different correlations among the couplings. The models considered in this work are the following:
\begin{enumerate}

\item \emph{General model:} The first model makes no additional assumptions regarding the coefficients $\c X$, which are all allowed and free to vary independently. This represents the least restrictive choice possible in our given EFT framework, and includes a total of 15 coefficients in addition to the mass of the DM.

\item \emph{Flavour-blind:} In this model, all operators involving fermions with the same gauge quantum numbers, e.g., all left-handed (LH) quarks, are assumed to also have the same couplings to DM and therefore have the same coefficients in the effective lagrangian. We are then left with five different coupling parameters which are related to those of the general model as
\begin{gather}
\cL e = \cL\mu = \cL\tau \equiv \cL{}, \quad \cQ 1 = \cQ 2 = \cQ 3 \equiv \cQ{}, \nonumber \\ \cu = \cc = \ct, \quad
\cd = \cs = \cb, \quad \ce = \cmu = \ctau.
\end{gather}
For the latter three groups of coefficients, we will generally use the notation for the coefficient of the first generation to refer to it within this model.

\item \emph{Family-oriented:} In this model, we make the assumption that the effective couplings with DM are equal for all particles belonging to the same generation. This can be considered a quite crude proxy to flavour theories were the successive SM families are characterized by hierarchical couplings or charges in order to explain the observed mass hierarchy between them, following the philosophy of the Froggatt-Nielsen mechanism\footnote{See \cite{Calibbi:2015sfa} for a realisation in the context of dark matter.} \cite{Froggatt:1978nt,Leurer:1992wg,Leurer:1993gy,Ibanez:1994ig,Binetruy:1994ru}. This leaves only three independent operator coefficients, 
\begin{gather}
\cL e = \ce = \cQ 1 = \cu = \cd \equiv \c 1, \quad \cL\mu = \cmu = \cQ 2 = \cc = \cs \equiv \c 2, \nonumber \\
\cL\tau = \ctau = \cQ 3 = \ct = \cb \equiv \c 3.
\end{gather}
This is the model with the fewest number of free parameters which we will consider and, as such, the correlations among the couplings will allow to obtain strong constraints, particularly for the coupling $c_1$.

\item \emph{Leptophobic:} This model assumes that DM does not have significant interactions with any of the leptons. As a result, all of the coefficients associated to operators involving leptons are set to zero, i.e.,
\begin{equation}
 \cL{\ell} = c_{\ell R} = 0,
\end{equation}
with $\ell = e, \mu, \tau$. At the same time, no restrictions are imposed on the operators involving quarks and the model therefore contains nine free parameters.

\item \emph{Leptophilic:} In direct contrast to the leptophobic model, this model instead considers the situation where the only relevant DM interactions with the visible sector are those involving leptons. In this model, we therefore set all of the coefficients for operators involving quark fields to zero, \ie ,
\begin{equation}
\cQ{i} = c_{u_i R} = c_{d_i R} = 0 \, ,
\end{equation}
where $i$ indicates the generation. In this situation, we are left with six independent coupling parameters for the DM interactions with leptons.
\end{enumerate}

We wish to stress that these models are only meant to be \emph{phenomenological tools} to assess our present global knowledge of how well constrained the DM couplings to the SM fermions are when not assumed to be universal or related through any particular UV completion. We will thus allow all couplings to vary freely when fitting the present data without questioning the apparent \emph{naturalness} (or lack thereof) of the preferred regions. In particular, as we will see in Sec.~\ref{sec:results}, most models prefer $2\cQ1 + \cu + \cd \simeq 0$ so as to avoid the very stringent limits coming from direct detection searches. Thus, even if these cancellations may seem unnatural, we will not avoid them by adding artificial ``naturalness priors'' to guide the Monte Carlo in any way, in the spirit of allowing the Monte Carlo to choose the points in parameter space which provide the best fit to the data. Symmetry arguments could perhaps be invoked in a particular UV completion when trying to reproduce the best fit found in the effective description in a natural way.  

Moreover, these working models are \emph{not} intended to be self-consistent low energy descriptions of any particular UV completion. In fact, due to the limited amount of operators considered, we are not allowed to take our effective theory description beyond tree level processes. Indeed, radiative corrections could induce other operators with interesting phenomenology.\footnote{For recent works in an EFT framework including one loop processes, see e.g. Refs.~\cite{Crivellin:2014qxa,D'Eramo:2014aba}} For example, in the leptophilic model the lepton legs could be closed in a loop and, through the emission of a virtual photon, induce a coupling to first generation quarks. Therefore, the strong bounds from direct searches would also apply to these models and put a constraint on the lepton couplings (see Ref.~\cite{Kopp:2009et}). In a similar fashion, other signals at the LHC such as dijet or dileptons could be generated from loop-induced 4-SM-fermion operators. Furthermore, when moving from the flavour basis to the mass basis, the DM couplings to the SM doublets will induce flavour-changing operators that, when the DM legs are closed in a loop, could contribute to the oscillations of neutral kaons or other FCNC processes, for which stringent experimental constraints would apply. In full consistency, these operators should have been included from the start, since they are compatible with the particle content and symmetries of the theory and they are required to renormalize the divergences stemming from the loop contributions. However, when doing so, new unbounded coefficients are introduced and, in order to avoid these stringent limits, the fit will prefer the points in the parameter space where these new coefficients exactly cancel the loop-induced contributions. Thus we will simply not consider these loop-induced constraints in our list of observables, since in our approach they would not imply additional constraints on our parameter space unless particular relations between the couplings are invoked.

\section{Considered constraints on DM}
\label{sec:constraints}

In our numerical analysis, the DM contributions to the different DM observables are computed with \micrO~\cite{Belanger:2013oya}. These are then compared to the current experimental bounds. A chi-square function is then computed for each observable independently. The total $\chi^2$ is obtained as the sum of all separate contributions

\begin{equation}
\chi^2(\c{},m_\chi) = \sum_i \chi_i^2(\c{},m_\chi),
\end{equation}
where $i$ is an index which runs over the observables and $\c{}$ is a vector containing the coupling coefficients of the model being tested.

The contribution to the chi-square from each experiment is computed assuming that the limit on the physical quantity being bounded (\textit{i.e.,} relic abundance, the thermally averaged cross section, etc) by the experimental collaborations behaves like a gaussian. The chi-square can be expressed as a function of the number of events as:
\be 
\chi^2 = \left( \frac{N^{th}}{\sqrt{N^{th} + N_{bg} + \Delta N_{sys}}} \right)^2\, ,
\label{eq:chi2-general}
\ee
where $N^{th}$ and $N_{bg}$ correspond to the number of signal and background events, respectively, and $\Delta N_{sys}$ refers to the systematic uncertainty on the number of events. Notice that, given the unfortunate lack of a positive DM signal in any of the observables described below with the exception of the relic abundance constraint, in this expression it has been assumed that the observed number of events corresponds with the expected background. In some cases, small upward or downward fluctuations over that background will be present and thus small deviations with respect to the scaling derived below can take place. However, notice that we will normalize this scaling with the data presented by each experimental collaboration itself. Thus, our $\chi^2$ function will always reproduce correctly the result obtained by the collaboration at the confidence level at which it is reported (typically 90 or 95\%) and the deviations should be small for the purpose of our exploration as long as CL close to these limits are investigated.   

Experimental collaborations present their results as bounds on a given quantity at a certain confidence level (CL). This can be translated into a certain number of events $N^{exp}$ allowed at that CL. For a bound at 68\% CL, for instance, we get\footnote{When the experimental result is quoted at a different CL, it can be simply rescaled to 68\% CL within a gaussian approximation. }:
\be 
\chi^2_{68\%} = 1 = \left( \frac{N^{exp}}{\sqrt{N^{exp} + N_{bg} + \Delta N_{sys}}} \right)^2 \, .
\label{eq:chi2-1}
\ee

By taking the ratio between Eqs.~\eqref{eq:chi2-general} and~\eqref{eq:chi2-1}, we get the following expression for the chi-square:
\be 
\chi^2 = \left( \frac{N^{th}}{N^{exp}} \right)^2 \frac{N^{exp} + N_{bg} + \Delta N_{sys}}{N^{th} + N_{bg} + \Delta N_{sys}} \, .
\ee

The expression above can be further simplified in the two following limits:
\begin{itemize}
 \item Statistically-dominated: in this case, the number of background events and the systematic errors can be neglected, and we can therefore write the chi-square contribution as:
 \be 
 \chi^2 \simeq \frac{N^{th}}{N^{exp}} \, . 
 \label{eq:chi2-stat} 
 \ee

 \item Background/Systematics-dominated: in this case, the signal events can be neglected with respect to the background and/or the systematic error, and the chi-square simplifies to:
 \be 
 \chi^2 \simeq \left( \frac{N^{th}}{N^{exp}} \right)^2 \,  . 
 \label{eq:chi2-sys} 
 \ee
\end{itemize}

By making use of Eqs.~\eqref{eq:chi2-stat} or~\eqref{eq:chi2-sys}, all the finer details of the detector response cancel in the ratio. We will therefore use the experimental bounds on the different observables reported by the collaborations instead of the number of events (which are not always publicly available).

All the experiments considered in this work fall in the background/systematics-dominated case, and therefore we will apply Eq.~\eqref{eq:chi2-sys}. The only exception to this will be the case of dwarf galaxies studied by the Fermi-LAT collaboration~\cite{Ackermann:2015zua}. These are dark-matter--dominated objects and constitute extremely clean probes to test for dark matter interactions with SM particles. We have explicitly checked the behaviour of the chi-square for a subset of the observed dwarf galaxies, using publicly available data from the collaboration from Ref.~\cite{Ackermann:2015zua}, and we find that the behaviour is in between the two extreme cases identified above. We therefore opt for the most conservative approach in this case, which corresponds to Eq.~\eqref{eq:chi2-stat}, when using the limit provided by the collaboration (which has been obtained using the full data sample).

In the following, we describe the full set of observables sensitive to interactions between the DM and the SM fermions which have been included in our simulations.

\subsection*{Relic abundance} 

The abundance of DM in the Universe is well-known from the PLANCK measurements~\cite{Ade:2015xua}. At present, its central value is 
\begin{equation}
\Omega_\chi^{\rm exp} h^2 = 0.1187, 
\end{equation}
with an error of $\sigma_\Omega = 0.0012$ at $1\sigma$. We will assume that our DM fermion $\chi$ constitutes all of the observed relic density and that it is produced in the early Universe through thermal freeze-out. The predicted relic density $\Omega_\chi^{\rm th}(\c{},m_\chi)$ is computed for each set of values for the model parameters, and the corresponding contribution to the chi-square function is given by
\begin{equation}
\chi^2_{\Omega}(\c{},m_\chi) = \left( \frac{\Omega_\chi^{\rm th}(\c{},m_\chi) h^2  - \Omega_\chi^{\rm exp}h^2 }{\sigma_\Omega}\right)^2.
\end{equation}
In general, within the thermal freeze-out scenario, the relic abundance of DM is mainly governed by the thermally averaged total DM annihilation cross section
\begin{equation}
\langle\sigma v\rangle = \sum_i w_i \langle\sigma v\rangle_i \propto m_\chi^2 \Sigma_C^2 ,
\quad
\mathrm{where}
\quad
\Sigma_C = \sqrt{\sum_i w_i c^2_i} \, .
\label{eq:Sum}
\end{equation}
Here, $i$ runs over all fermion fields contributing to the annihilation cross section, and $w_i$ is a weight associated to the dimension of the SM gauge group representation of the corresponding fermion field. As we will show in the results section, this combination is subject to very strong constraints and dictates the possible ranges in which the coefficients may lie. In particular, due to the finite and non-zero relic abundance and the assumption of thermal freeze-out, this implies that at least one of the coefficients must be non-zero and that none of them can be too large. It should be kept in mind that, alternatively, under the assumption of freeze-in the correct relic abundance could be generated via extremely small couplings to all SM fermions~\cite{Hall:2009bx}.  

\subsection*{Direct detection} 

The contribution to the chi-square from direct detection experiments is computed assuming that the limit on the spin-independent cross section given by the experimental collaborations behaves like a gaussian and is background-dominated (see Eq.~\eqref{eq:chi2-sys}). Therefore, the contribution to the chi-square coming from each direct detection experiment can be computed as
\begin{equation}
\chi^2_{DD}(\c{},m_\chi) = \left( \frac{ \sigma^{\rm th}_{DD}}{\sigma^{\rm exp}_{DD}}\right)^2 ,
\end{equation}
where we have expressed the limit $\sigma^{\rm exp}_{DD}$ at the 68\% CL (after rescaling from the 90\% quoted by the collaborations). Note that, for $\sigma^{\rm th}_{DD} = \sigma^{\rm exp}_{DD}$, the appropriate value of the chi-square function and the experimental limit are recovered.

For $\sigma^{\rm exp}_{DD}$, we implement direct detection constraints from LUX~\cite{Akerib:2013tjd} and, so as to have an independent target material, EDELWEISS-II~\cite{Ahmed:2011gh}. Our setup generally leads to a spin-independent contribution (except when $c_{Q1} = -c_{uR} = -c_{dR}$) which will be constrained by the LUX and EDELWEISS-II results. For our set of operators, the spin-dependent contribution cancels out~\cite{Cheung:2012gi}. The spin-independent dark-matter--nucleus coupling $\c{N}$ will relate to the quark couplings as
\begin{equation}
 \c{N} = \frac{1}{2A}[3A\, \cQ1 + (A+Z)\cu + (2A-Z)\cd],
 \label{eq:degeneracy}
\end{equation}
where $A$ is the mass number and $Z$ the atomic number of the target nucleus. With this dependence on the couplings, there is a possible degeneracy in the DM--quark couplings for which the cross-section vanishes. For $Z \simeq A/2$, this degeneracy occurs for $2\cQ1 + \cu + \cd \simeq 0$. Thus, direct detection experiments will only allow large couplings to the first generation quarks if this particular relation is fulfilled. As we will show in Sec.~\ref{sec:results}, this is clearly seen from the numerical results of our simulations.

\subsection*{Cosmic Microwave Background (CMB)}  

DM annihilations result in an injection of power into the Intergalactic Medium (IGM) per unit volume equal to \cite{Chen:2003gz,Padmanabhan:2005es}
\be
\left(\frac{{\rm d}E}{{\rm d}t{\rm d}V}\right)_{\rm inj} = (1+z)^6 (\Omega_{\chi,0}\rho_{c,0})^2 \zeta\df{\langle\sigma v\rangle}{m_\chi},
\ee
where $z$ is the redshift, $\Omega_{\chi,0} (\rho_{c,0})$ is the DM abundance (critical density) today, $\langle\sigma v\rangle$ is the thermally averaged annihilation cross section, and the statistical factor $\zeta=1/2$ corresponds to DM being a Dirac particle.
On the other hand, CMB probes such as the WMAP and Planck satellites can set limits on the deposited power into the IGM around the CMB epoch ($z\sim1000$), which is related to the injection power as \cite{Slatyer:2009yq,Slatyer:2012yq,Lopez-Honorez:2013lcm}.
\be
\left(\frac{{\rm d}E}{{\rm d}t{\rm d}V}\right)_{\rm dep} =  f_j(z, m_\chi)\left(\frac{{\rm d}E}{{\rm d}t{\rm d}V}\right)_{\rm inj} , 
\ee
where the efficiency function $f_j$ depends on the DM annihilation channel, $\chi\bar\chi\to p_jp_j$. For this reason, experiments usually quote their limits through the quantity 
\be
p_{\rm ann} = \sum_j f_j \df{\langle\sigma v\rangle_j}{2 m_\chi}~,
\ee
where $\langle\sigma v\rangle_j$ is the thermally averaged partial annihilation cross section into particles $p_j$.
This quantity has been constrained by Planck+lensing data at the 95\% CL~\cite{Ade:2015xua}, here we rescale that bound to the $68 \%$ CL obtaining: $p^{\rm exp}_{\rm ann} < 1.7\times 10^{-28}$~cm$^3$s$^{-1}$GeV$^{-1}$ so as to build our $\chi^2$ function . 
In order to implement these bounds we use the tabulated values of $f_j$ from Ref.~\cite{Slatyer:2009yq}, to compute  $p^{\rm th}_{\rm ann}$, and compare it with the experimental constraints. The contribution added to the total $\chi^2$ is given by
\begin{equation}
\chi^2_{\rm CMB}(\c{},m_\chi) = \left( \frac{p_{\rm ann}^{\rm th}}{p_{\rm ann}^{\rm exp}}\right)^2.
\end{equation}

\subsection*{Positron fraction} 
The measurements from AMS02 on the positron fraction~\cite{Aguilar:2013qda} are used to
derive upper bounds on DM annihilation cross sections. The contribution to electron and positron fluxes $\Phi_{e^\pm,{\rm DM}}$ from DM annihilation are computed using \micrO. The contributions coming from astrophysical sources, on the other hand, are parameterized by 
\begin{equation}
\Phi_{e^-,{\rm bg}}(E) = C_{1} E^{-\gamma_{1}} +
  C_2 E^{-\gamma_2}, ~~~ \Phi_{e^+,{\rm bg}}(E) = C_{e^+}
  E^{-\gamma_{e^+}} + C_s E^{-\gamma_s} e^{-E/E_s}
  \label{eq:posiflux}
\end{equation}
as in Ref.~\cite{Ibarra:2013zia},
where the last term in the positron flux represents a point source with a hard spectrum cut $E_s$, while the other
terms model the diffuse background. Both DM and background fluxes are affected by the
solar modulation, which can be explicitly taken into account by computing the flux at the top of the
atmosphere ($\oplus$) under the force field approximation:
\begin{equation}
  \Phi_{e^\pm}^{\oplus} (E) = \df{E^2}{( E + \phi^\pm )^2} \Phi_{e^\pm}( E + \phi^\pm )
  \label{eq:solarmod}
\end{equation}
where $\phi^\pm$ refer to the parameters accounting for the
modulation. The differential positron
fraction on top of the atmosphere is then given by
\begin{equation} 
{\cal  F} =\df{\Phi^{\oplus}_{e^+,{\rm bg}} + \Phi^{\oplus}_{e^+,{\rm DM}}}
{\Phi^{\oplus}_{e^+,{\rm bg}} + \Phi^{\oplus}_{e^+,{\rm DM}}+\Phi^{\oplus}_{e^-,{\rm bg}} +
  \Phi^{\oplus}_{e^-,{\rm DM}}}.
  \label{eq:posifraction}
\end{equation}
The $\chi^2$ function is built by binning the predicted positron fraction
and comparing to the experimentally measured values,
\begin{equation} 
\chi^2_{{\rm AMS}}(\c{},m_\chi) =
\sum_j^{\rm bins}\left(\df{{\cal F}_{j}-{\cal F}_{{\rm exp},j}}{\sigma_{{\cal F},j}}\right)^2,
\end{equation}
where the uncertainties in each bin $\sigma_{{\cal F},j}$ are taken
directly from Ref.~\cite{Aguilar:2013qda} by adding the statistical and systematic errors
in quadrature. 

We have checked that allowing the electron flux to vary does not affect the fit in any
substantial way and have therefore fixed the parameters for this flux to the values which provide the best-fit to Fermi electron flux data~\cite{Ackermann:2010ij}. In particular, the value of the solar modulation parameter for the electron flux which gives a best-fit to the Fermi data is found to be equal to zero. Moreover, the solar modulation parameter for positrons $\phi^+$ was found to have a negligible impact on the fit, while creating some numerical degeneracies. Therefore, we choose to fix this parameter to zero as well during the fit, as described in detail in App.~\ref{sec:app}. However, it should be noted that the solar modulation only affects the low energy part of the spectrum and therefore our analysis of AMS02 is expected to be accurate for $m_\chi\gtrsim 10~{\rm  GeV}$.

In order to take into account the lack of knowledge on the
astrophysical backgrounds, we profile over the positron flux parameters in Eq.~(\ref{eq:posiflux}) in our simulations,\footnote{The minimisation procedure was performed with the {\tt GSL} libraries \cite{GSL}.} without imposing any priors on them. This is done for each set of parameters $\{c\}$ independently, as explained in detail in App.~\ref{sec:app}.

\subsection*{Gamma rays from dwarfs} 

Dwarf spheroidal satellite galaxies (dSphs), being DM-dominated objects, constitute very clean laboratories to test DM interactions. A search for $\gamma$-rays coming from Milky Way dSphs has been performed by the Fermi-LAT collaboration~\cite{Ackermann:2015zua}. We make use of the present $90 \%$ CL bounds (rescaled to the $68 \%$ CL) on the thermally averaged cross section for DM annihilation $\langle\sigma v\rangle^{\rm exp}_j$ from Ref.~\cite{Ackermann:2015zua}, which are derived under the assumption of 100\% DM annihilation into some specific channel $\chi\bar\chi\to p_jp_j$. However, in our EFT approach, DM may annihilate to all fermionic channels with different branching fractions $B_j$, which will depend on the set of couplings for a given model $\{ c \} $, and each particular channel $j$ will contribute to $\gamma$-ray emission with a different weight, given its different subsequent decay chains. This weight is inversely proportional to the final experimental bound that can be derived for that particular channel $\langle\sigma v\rangle^{\rm exp}_j$. We may therefore recast the experimental bound on the total cross section as
\be
\langle \sigma v\rangle^{\rm exp}(c, m_\chi) =\left( \sum_{j} \df{B_j}{\langle\sigma v\rangle^{\rm exp}_{j}(m_\chi)}\right)^{-1} .
\ee
As already mentioned, the contribution to the chi-square coming from dwarf galaxy measurements is not completely dominated by either background/systematics nor by statistics. Therefore, we conservatively assume that it is the latter which dominates the measurement in this case. The contribution to the $\chi^2$ is thus given by
\be
\chi^2_{\rm dSphs}(\c{},m_\chi) = \df{\langle\sigma v\rangle^{\rm th}}{\langle\sigma v\rangle^{\rm exp}} ~,
\ee
where $\langle\sigma v\rangle^{\rm th}$ is the total annihilation cross section today as predicted by the model being tested with couplings $\c{}$ and DM mass $m_\chi$. 

\subsection*{Monojets at LHC} 

Here we take the most recent 95\% CL LHC results (rescaled to the $68 \%$ CL) from monojet+$E^T_{\rm miss}$ analyses \cite{Aad:2015zva}, applied to the effective vector interaction operator $\bar{q} \gamma^\mu q \bar{\chi} \gamma_\mu \chi$, where a universal coupling to up and down-type quarks of both chiralities was assumed.\footnote{In view of the possible issues regarding the validity of the EFT approach in collider searches~\cite{Busoni:2013lha}, we consider the limits from~\cite{Aad:2015zva} which respect the validity criteria of EFT. In the next LHC run, the issues with the validity of the EFT will be even more relevant, and therefore an analysis in terms of Simplified Models \cite{Buchmueller:2014yoa,Abdallah:2014hon,Buckley:2014fba,Abdallah:2015ter,Abercrombie:2015wmb} might be preferable.}
Notice that, if opposite couplings are assumed for the different chiralities, the same bound is recovered~\cite{Goodman:2010ku}. Indeed, any interference effect between the different effective operators will be chirality suppressed. Thus, it is safe to assume that the total contribution can be obtained as the incoherent sum of the individual ones. Since a full collider simulation is beyond of the scope of this work, we recast the existing limits on the coefficient $C_{\rm LHC}^{\rm exp}$ of the effective operator as a function of the DM mass. Since the constraint mainly stems from events with large $Q^2$, for which the valence quarks dominate the parton distribution function of the proton, we assume that only the first generation quarks participate. Thus, the contribution to the $\chi^2$ function will be given by
\be
\chi^2_{\rm LHC}(\c{},m_\chi) =\left(\df{(C_{\rm LHC}^{\rm th})^2}{6 (C_{\rm LHC}^{\rm exp})^2}\right)^2~,
\ee
where $(C_{\rm LHC}^{\rm th})^2 = 3 \cQ{1}^2 +2\cu^2 + \cd^2$.

\subsection*{Mono-photons at LEP}  Similarly to the monojet case, the 95\% CL constraint (rescaled to the $68 \%$ CL) on the coefficient of the $\bar{e} \gamma^\mu e \bar{\chi} \gamma_\mu \chi$ operator, $C_{\rm LEP}^{\rm th}$, by LEP mono-photon+$E^T_{\rm miss}$ searches is considered \cite{Fox:2011fx}. The contribution to the $\chi^2$ is thus:
\be
\chi^2_{\rm LEP}(\c{},m_\chi) = \left(\df{ (C_{\rm LEP}^{\rm th})^2}{2 (C_{\rm LEP}^{\rm exp})^2}\right)^2~,
\ee
where $(C_{\rm LEP}^{\rm th})^2 = \cL{e}^2 +\ce^2 $.

\section{Numerical details}
\label{sec:scan}

All the phenomenological models described in Sec.~\ref{sec:setups} have been implemented in {\tt CalcHEP}~\cite{Belyaev:2012qa} using {\tt FeynRules}~\cite{Alloul:2013bka}. The result is then fed into {\tt micrOmegas}~\cite{Belanger:2013oya} for the computation of the DM observables.

With the chi-square function implemented as described above, the parameter spaces of the different models need to be efficiently explored in order to derive constraints. While a simple grid scan may be viable in the case of the family-oriented model (with only three parameters), it quickly becomes prohibitively expensive for models with a larger number of free parameters. In particular, the general model with 15 free parameters would be unfeasible to scan in this fashion. Moreover, we expect the parameter space to be affected by several degeneracies. These further decrease the effectiveness of a grid scan and require the grid spacing to be very small in order to find the global minimum. For these reasons, we have opted to perform our scans using the nested sampling~\cite{Skilling:2004} Monte Carlo algorithm implemented in the {\tt MultiNest} software~\cite{Feroz:2008xx}. This method is particularly designed for handling parameter space degeneracies as well as for preferentially scanning the regions of parameter space where the likelihood function $\mathcal{L} = \exp(-\chi^2/2)$ is large. Although the {\tt MultiNest} software was initially designed for computing Bayesian evidence, it produces a sample of the parameter space with the corresponding likelihood values as a by-product. In this paper, we take a purely frequentist approach and only consider the $\chi^2$ values obtained from the likelihood. Thus, {\tt MultiNest} is used only for its capability of efficiently sampling the regions with relatively low $\chi^2$ values.

In our {\tt MultiNest} simulations, the number of live points was set to 750, the tolerance to 0.2 and the efficiency to 0.9. Constant efficiency mode was not used. The range of parameters scanned was adapted to the values of the coefficients required to reproduce the correct relic abundance, for each of the dark matter masses considered. The scan was done in linear scale over both positive and negative values for the coefficients accompanying each of the operators. The number of distinct samples obtained ranges between $10^5$ and $5\times 10^5$, depending on the particular model and DM mass considered.

\section{Results}
\label{sec:results}

This section describes the results obtained from a global fit using
the models described in Sec.~\ref{sec:setups}. All constraints considered in
Sec.~\ref{sec:constraints} have been included in our simulations.

Note that the dimensionful couplings $c_j$ will depend on the effective theory mass scale $\Lambda$ as $\Lambda^{-2}$. In this section, we will use units of TeV$^{-2}$ for the couplings $c_j$ and the rough limit  $\Lambda^2 \lesssim 1/c_j$ may be used to assess the range of validity of the EFT assumption.

Using the $\chi^2$ function built as described in Sec.~\ref{sec:constraints}, we have derived two main types of constraints on the parameters of the model:
\begin{itemize}
\item[i)] For a fixed DM mass $m_\chi$, we scan the parameter space for all couplings in each model. This results in a sample of points in the parameter space, which is generally concentrated in the regions corresponding to the lowest values of the $\chi^2$ function. We use these to perform parameter estimation and determine the allowed regions for the different couplings within a given model, for given values of $m_\chi$. 
\item[ii)] The calculation described in (i) can be repeated for several values of the DM mass. By minimizing the global $\chi^2$ over all couplings $c$ in the model, the  resulting $\Delta \chi^2$,
\begin{equation}
\Delta \chi^2(m_\chi) = \min_{\c{}}\chi^2(\c{},m_\chi) - \min_{m_\chi,\c{}} \chi^2(\c{},m_\chi),
\label{eq:Deltachi2}
\end{equation}
allows parameter estimation of the DM mass $m_\chi$. By deriving confidence intervals around the global minimum of a given model, this will indicate whether there are any values of the DM mass which are disfavoured by current data.  Since the simulations are computationally rather expensive, they have been performed for a few values of the DM mass, namely $m_\chi = 10$, 50, 100, 200, 500 and 1000~GeV. Our minimization in the $m_\chi$ variable is therefore performed in a discrete fashion, using only this set of values. However, from $m_\chi \geq 200$~GeV on, we find essentially no change in the $\chi^2$ value. 
\end{itemize}

\begin{figure}
\centering
\includegraphics[width=0.6\textwidth,angle=0]{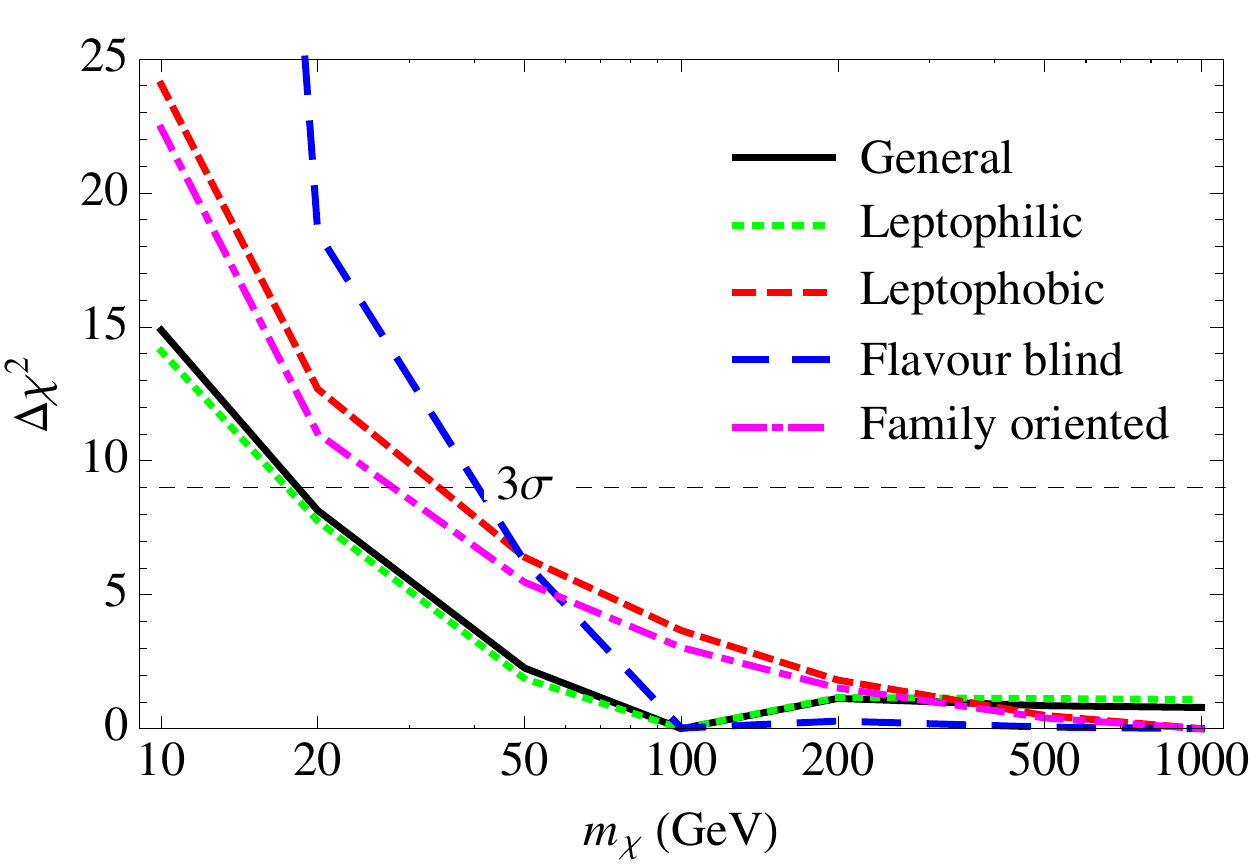}
\caption{\footnotesize{Value of the $\Delta \chi^2$ (see Eq.~\eqref{eq:Deltachi2}) obtained for the different models under consideration (indicated in the legend) as a function of the DM mass in GeV. For reference, the value of the $\Delta\chi^2$ corresponding to the $3\sigma$ limit (for 1 d.o.f.) is also shown by the dashed horizontal line.}}
\label{fig:chi2}
\end{figure}

\begin{figure}
\begin{tabular}{cc} 
\multirow{2}[2]{*}[60mm]{\hspace{1mm} \includegraphics[width=0.45\textwidth,angle=0]{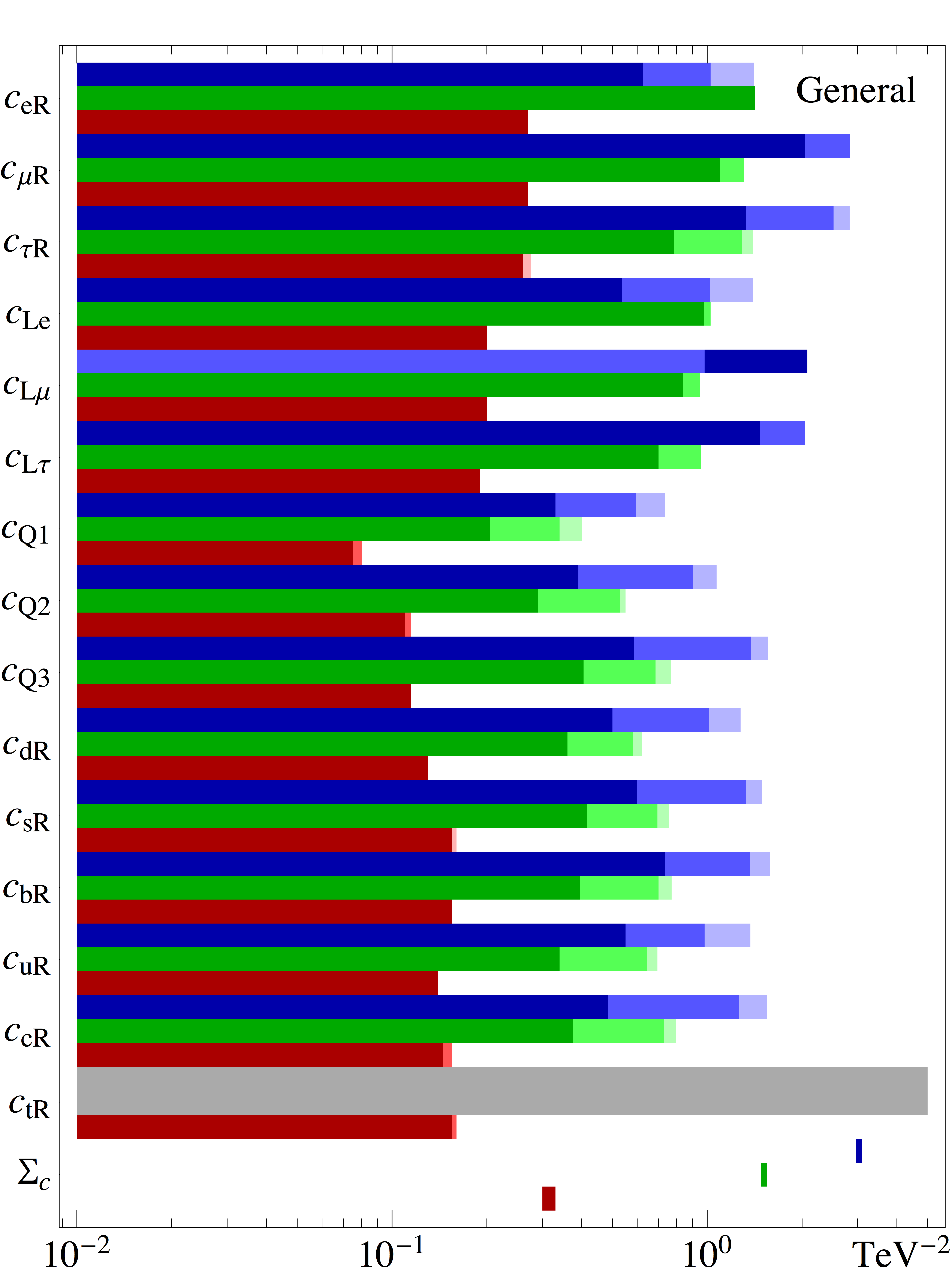} } & 
\hspace{5mm} \includegraphics[width=0.45\textwidth,angle=0]{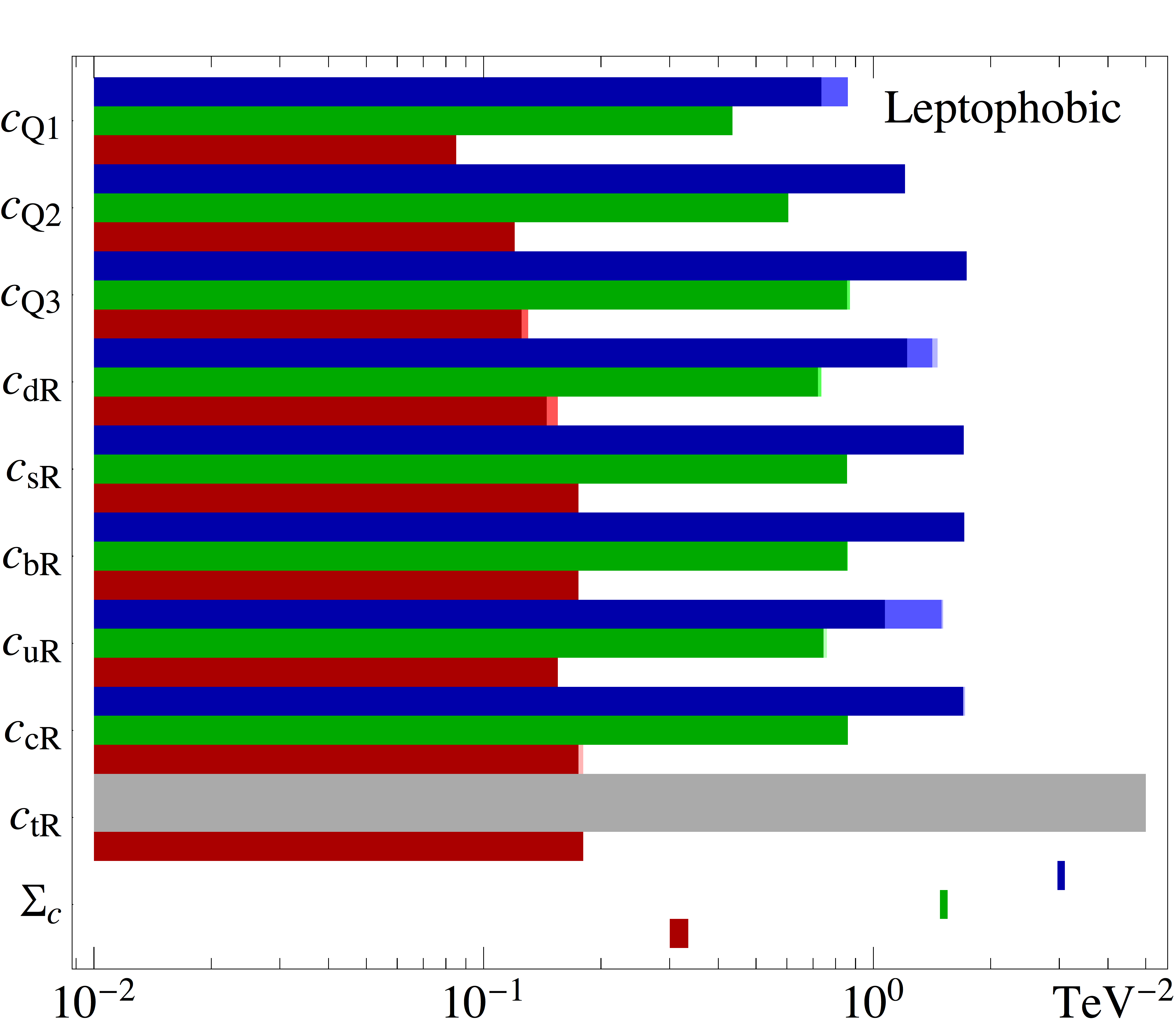}  \cr
& \hspace{-3.8mm } \includegraphics[width=0.505\textwidth,angle=0]{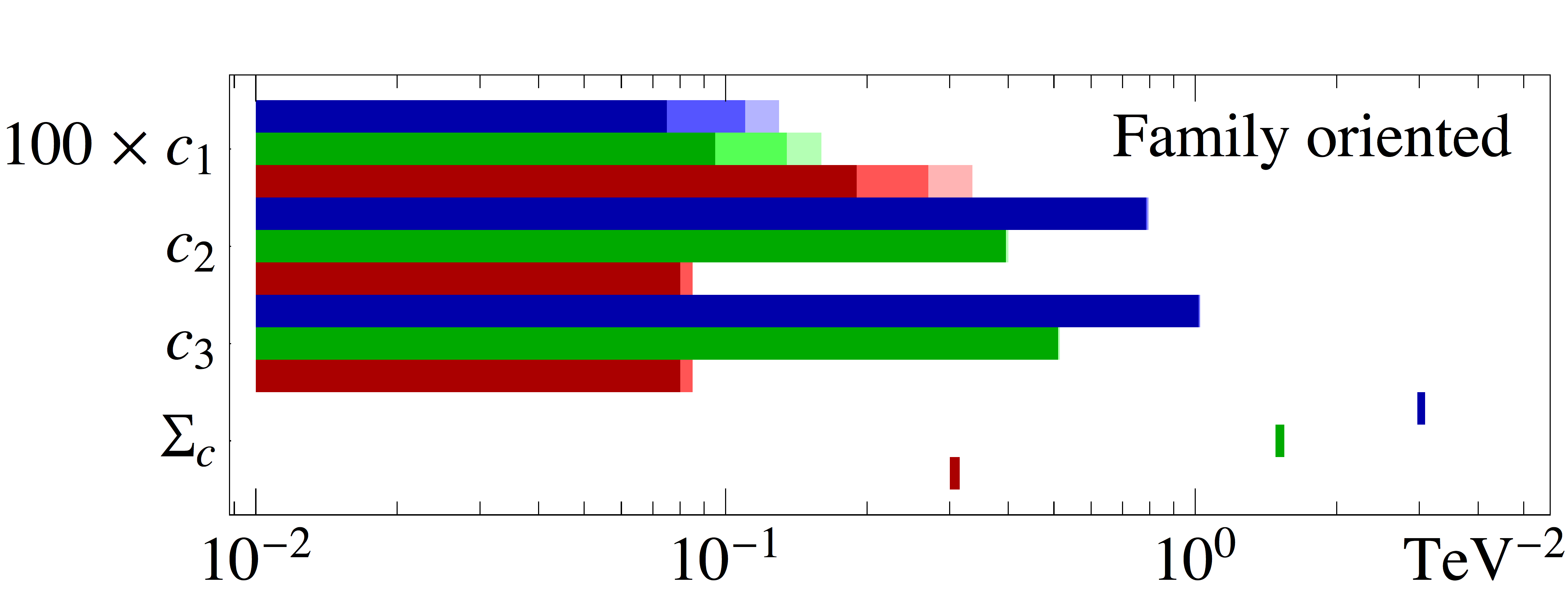} \\
\includegraphics[width=0.45\textwidth,angle=0]{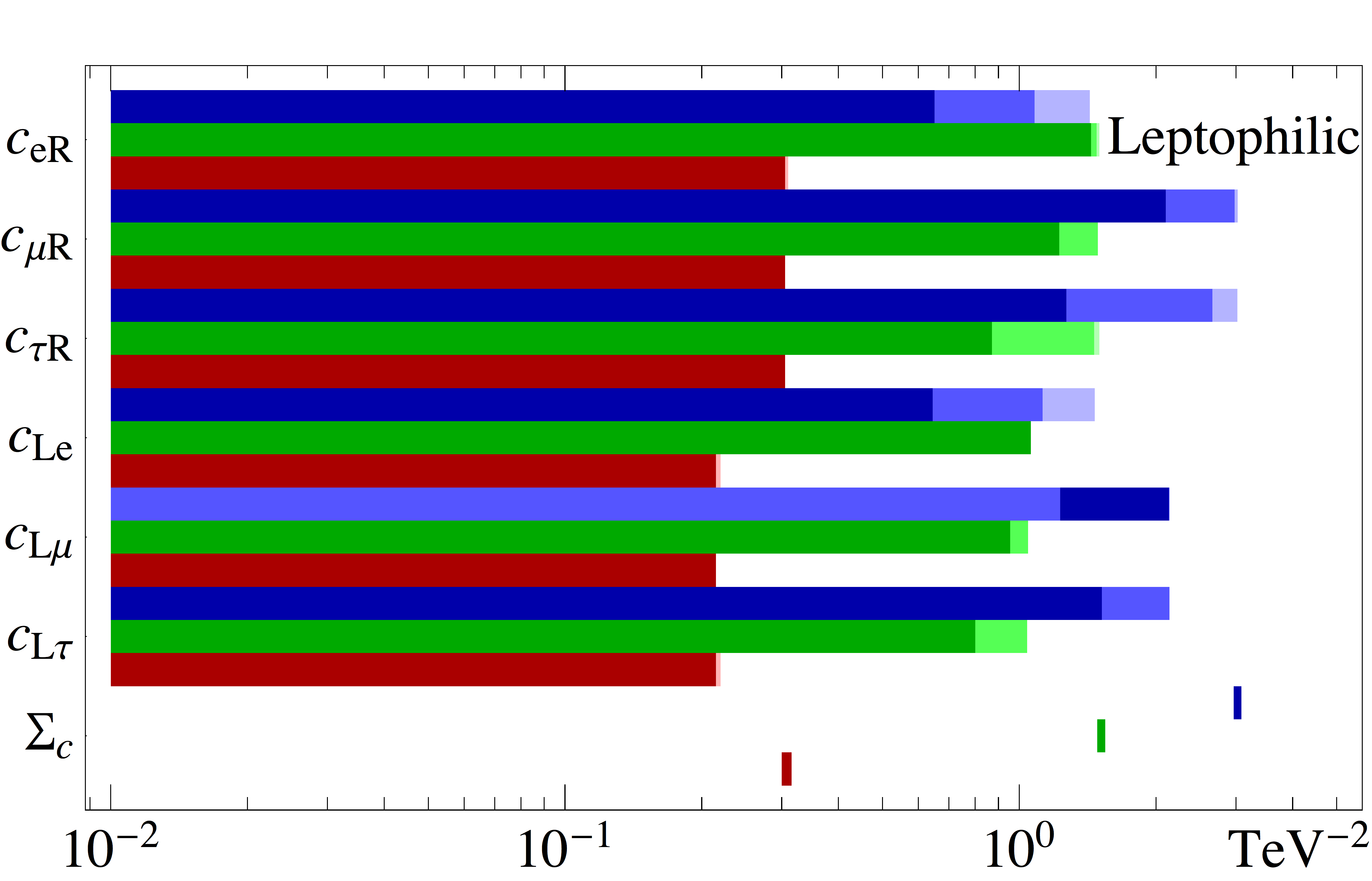} & 
\hspace{6mm} \raisebox{0.15\height}{\includegraphics[width=0.45\textwidth,angle=0]{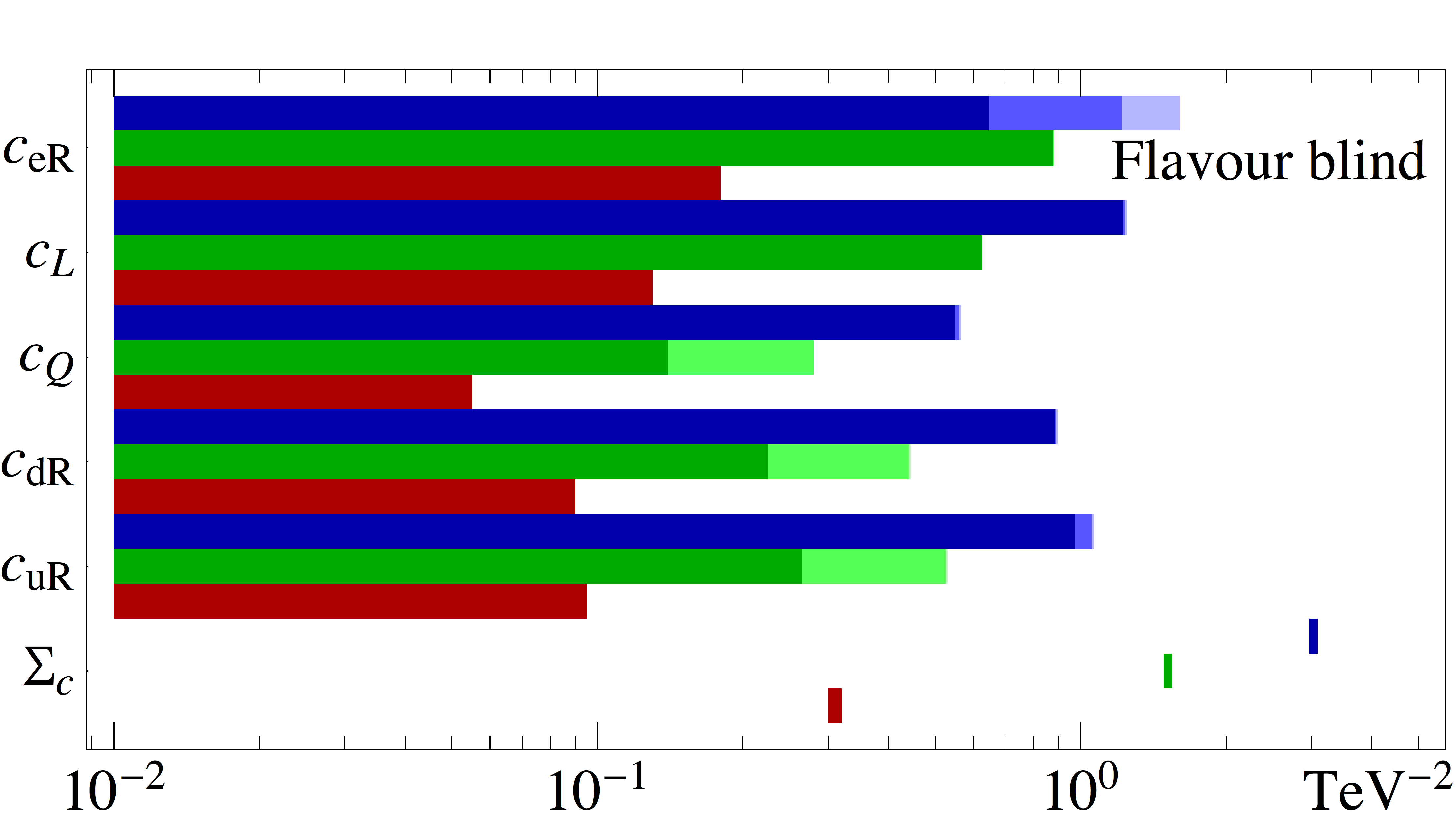} }  
\end{tabular}
\caption{\footnotesize{Allowed absolute values for the couplings associated to the different operators present in each of the models considered in this work. Dark, medium and light shaded areas correspond to the allowed regions at 1, 2 and $3\sigma$, respectively, for 1 d.o.f.. For each coupling, the results shown by the upper/blue, middle/green and lower/red bands correspond to DM masses of 50, 100 and 500~GeV, respectively. The gray bands indicate that the coupling is unconstrained since the process $ \chi \bar\chi \rightarrow t \bar{t} $ is not kinematically allowed for the considered value of the DM mass. }}
\label{fig:barcode}
\end{figure}

Our main results are summarized in Figs.~\ref{fig:chi2} and~\ref{fig:barcode}. Figure~\ref{fig:chi2} shows the value of the $\Delta \chi^2$ (see Eq.~\eqref{eq:Deltachi2}) obtained for the different models under consideration, as indicated in the legend, see item (ii) above. The dashed horizontal line shows the value of the $\Delta\chi^2$ corresponding to $3\sigma$ limit for 1 d.o.f.. As can be seen from this figure, DM masses around 40 GeV and below are disfavored at $3\sigma$ with respect to higher masses for most models under consideration. The only exceptions are the general and leptophilic models, for which only masses below 20~GeV are disfavored.

The fact that very light DM masses are generally disfavored can be understood from the complementarity between different data sets. As can be seen from Eq.~(\ref{eq:Sum}), the annihilation cross section is proportional to $m_\chi^2$, while the energy density scales as $m_\chi$. Thus, models with very light DM masses will require large couplings to the SM in order to satisfy relic density constraints. However, since there is no positive signal from DM in collider, direct or indirect detection data, a significant tension between the different data sets occurs, which eventually increases the minimum value of the $\chi^2$. The tension is stronger in models where DM either does not couple to leptons (leptophobic), or in models where the lepton couplings are related to others (\ie, flavour-blind  or family-oriented). For the general and leptophilic models, the tension is relaxed since the bounds which mainly constrain the couplings to leptons (LEP, CMB and AMS data) are not as strong as those constraining the quark couplings (direct detection, LHC and FermiLAT bounds). Finally, one can observe a slight preference in the fit for DM masses around 100~GeV for all models except for the leptophobic and family-oriented. This can be understood from the mild preference of AMS data for a DM signal  around this mass, since smaller masses are too strongly constrained, while heavier masses do not produce an appreciable signal. It should be stressed out, however, that this preference is far from being statistically significant. This feature is absent in the leptophobic and family-oriented models since the required fermion couplings for this signal to take place are either absent or constrained to be very small by other experimental bounds.

The second main result of this paper is shown in Fig.~\ref{fig:barcode}, where the allowed regions at 1, 2 and 3$\sigma$ (for 1 d.o.f.) are shown for all couplings associated to the effective operators and for the five models under consideration, see Sec.~\ref{sec:setups}. For each coupling, our results are shown for three different values of the DM mass, 10 GeV (upper/blue bands), 100~GeV (middle/green bands) and 500~GeV (lower/red bands). Each panel also shows the constraint imposed by relic density on the weighted sum of the squares of all couplings, \ie, $\Sigma_C$ as defined in Eq.~(\ref{eq:Sum}). 

The tension between different data sets can be understood from the results found in Fig.~\ref{fig:barcode}. As already explained, the constraint from relic density tends to bring the couplings to larger values as the DM mass is decreased. This is observed when comparing the upper/blue and lower/red bands, for all models under consideration, and in particular for the allowed values of $\Sigma_C$. One can clearly see how the tension between different data sets can lead to having only one coupling as the major contributor to the relic density for a given model. For instance, for the general and leptophilic models with $m_\chi = 50$~GeV, the relic density constraint is satisfied with a sizable coupling to the second family lepton doublet, and the preferred region for this coupling at $1\sigma$ does not include zero. This can be understood as follows. First, the bounds on couplings to quarks of all generations are very  strongly constrained by direct detection and Fermi-LAT experiments. Moreover,  on the leptonic side, the bounds on RH leptons from both AMS (electrons) and Fermi-LAT (taus) are very strong. In addition, LH couplings would also imply DM annihilations through muon neutrinos, so that the indirect searches are relaxed while the relic abundance constraint is more easily fulfilled. As such, the weakest global bound appears for the second lepton generation, which can accommodate the required relic abundance without being in tension with AMS or Fermi-LAT.

Nevertheless, a significant tension persists, as seen in Fig.~\ref{fig:chi2}. This is particularly the case for light DM masses, see, e.g., the bands corresponding to 50~GeV in Fig.~\ref{fig:barcode}. However, for 100 GeV mass (and heavier) the situation is different. For such large values, the indirect detection constraints are already weaker than the relic abundance one. Thus, since a coupling to LH fermions implies two annihilation channels, both contributing to decrease the DM density, we see that the LH couplings are more constrained than the corresponding RH couplings. Similarly, due to color multiplicity factors, couplings to quarks are more strongly constrained than those to leptons (in particular the LH ones).

Another interesting example is found in the panel for the flavour-blind model. Since the coupling to the muon is equal to that of the electron and tau (which are very strongly constrained) the relic abundance cannot be obtained out of annihilation to the second lepton doublet, as for the previous cases. For $m_\chi=50$ GeV, we see from the figure that the DM would prefer to annihilate to up-type RH quarks instead (the 1$\sigma$ region for this coupling is around the value of 1). However, this results in a strong tension with the Fermi-LAT data, and this mass is therefore disfavored at $\sim 3\sigma$ with respect to the best fit at higher mass (see Fig.\ref{fig:chi2}).

\begin{figure}
\centering
\includegraphics[width=0.45\textwidth,angle=0]{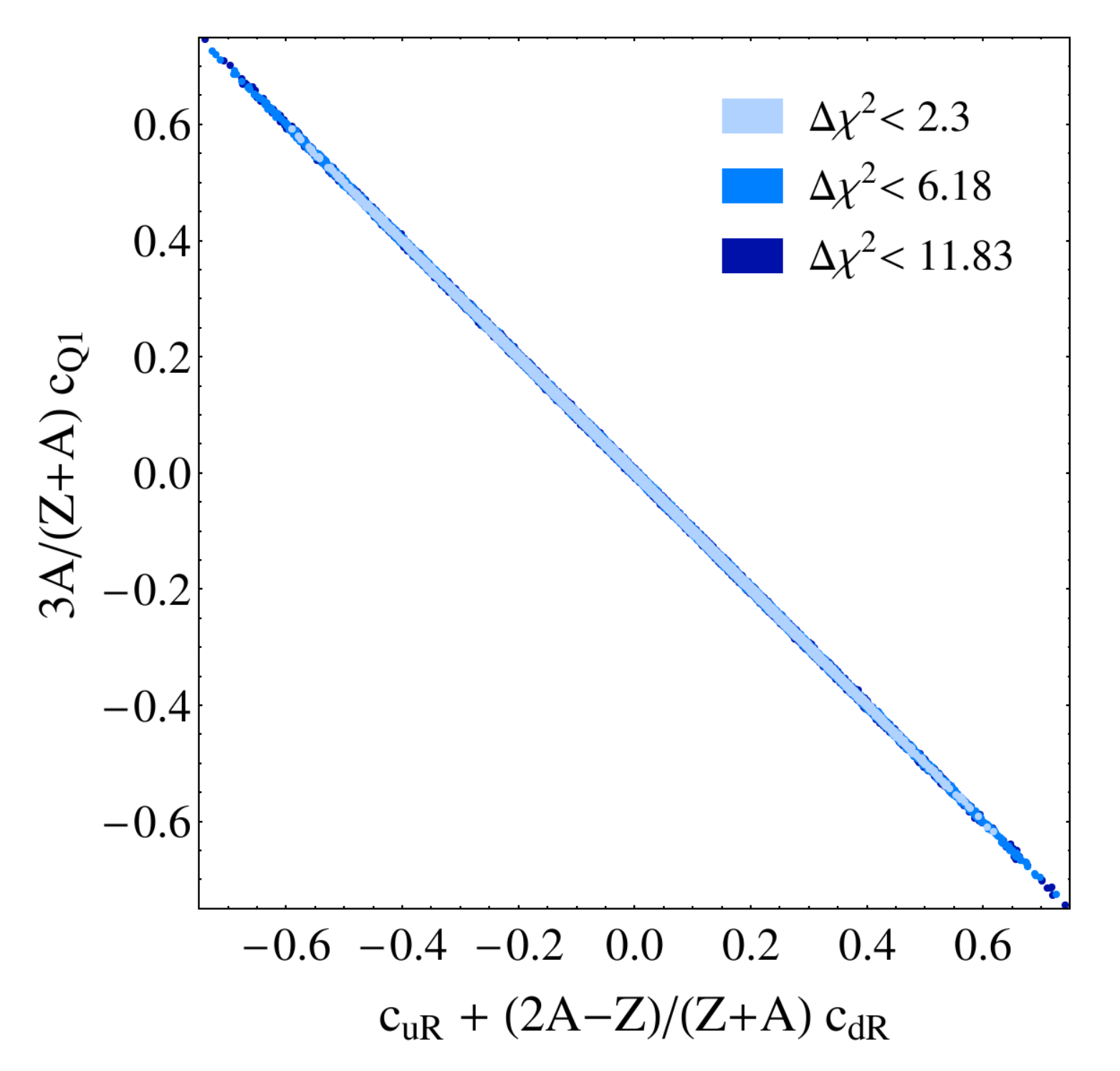}
\includegraphics[width=0.45\textwidth,angle=0]{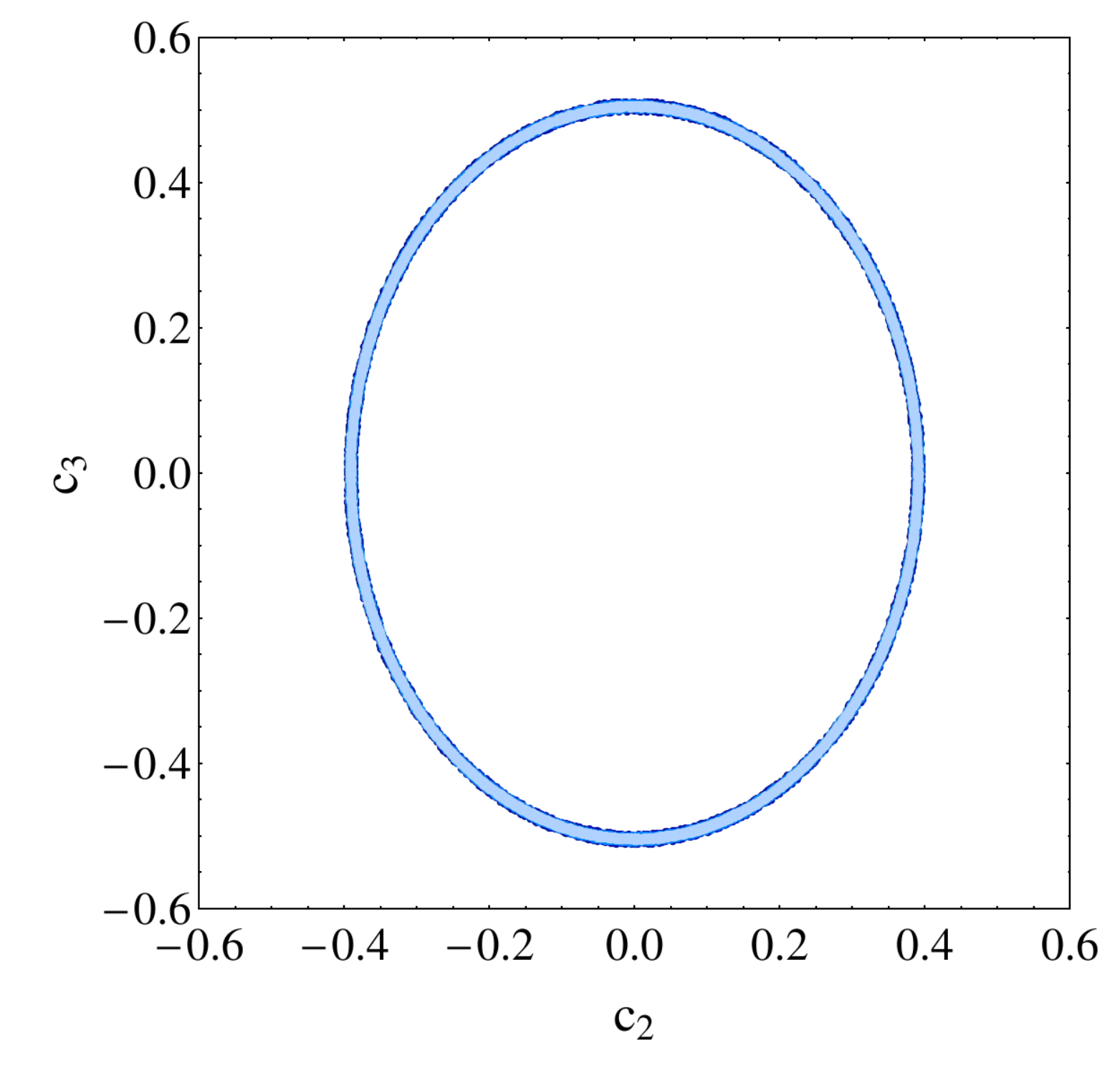}
\caption{\footnotesize{ Left: Allowed regions in the parameter space corresponding to the couplings to the first family of quarks for the general model. The degeneracy between $c_{Q1},c_{uR}$ and $c_{dR}$ is explicitly shown, see text for details. The weights of each coefficient are chosen so as to show explicitly the degeneracy condition in Eq.~\eqref{eq:degeneracy}. Right: Allowed regions for the couplings to the second and third family in the family oriented scenario. The point $c_2 = c_3 = 0$ is clearly disfavored due to the efficient combination of relic abundance and direct detection constraints, see text for details. In both panels, the different colors correspond to different confidence levels, indicated in the legend, and the DM mass is $ m_\chi = 100$~GeV.  }}
\label{fig:deg}
\end{figure} 

Finally, as was already mentioned in Sec.~\ref{sec:constraints}, we
have found two interesting degeneracies among different parameters in
the models considered in this work. The first one is due to the
constraints coming from direct detection experiments, which allows for
a degeneracy between the couplings to quarks. This is shown in the
left panel in Fig.~\ref{fig:deg}, where the allowed confidence regions
for the $\cQ 1$, $\cu$ and $\cd$ couplings are depicted, showing the degeneracy explicitly. For $m_\chi = 100$~GeV, the most stringent
bound on the effective operators involving the first generation of
quarks comes from direct detection experiments. Nevertheless, as can
be seen from this figure, the allowed regions extend to rather large
values of the couplings, as long as the relation $2.12\cQ 1 \simeq
-c_{uR} - 1.12 c_{dR} $ is satisfied, given the values of $Z$ and $A$
of the Xenon material used by LUX (see Sec.~\ref{sec:constraints} for
details). While the inclusion of EDELWEISS-II, with a different target material, could in principle lift this degeneracy, the difference in the ratio of protons and neutrons is not large enough for this task. The final limits found in Fig.~\ref{fig:deg} for the degeneracy line are rather stemming from the relic abundance constraint. 

The second degeneracy stems from the strong constraints from relic
abundance on $\Sigma_C$, which generally imply that all couplings must
lie on the surface of a hyperellipsoid. In particular, for the family
oriented scenario, this reduces to the ellipse displayed in the right
panel in Fig.~\ref{fig:deg}, where the allowed regions at 1, 2 and
3$\sigma$ are shown in the $c_2 - c_3$ plane for 2~d.o.f.. Given that
the strongest constraints from colliders and direct detection
experiments apply to the first generation, and in this model all
particles in the first generation have the same coupling, the
constraints on $c_1$ are very strong since the direct detection
degeneracy relation cannot be fulfilled. This is shown in
Fig.~\ref{fig:barcode}, where it can be seen that the coupling to the
first family is restricted to $ c_1 < 0.001/{\rm TeV}^2$ at $2\sigma$
for this model. This model only has two additional couplings, $c_2$
and $c_3$. Therefore, in order to satisfy relic density constraints,
either the coupling to the second or the third family (or both) have
to be different from zero. As a consequence, the allowed region is
shaped as an ellipse in the $c_2 - c_3$ plane.

A final remark regarding the validity of the EFT is in order. As
  can be seen from eq.~(\ref{eq:Sum}) and from Fig.~\ref{fig:barcode},
  the weighted sum of the coefficients required to obtain the correct
  thermal relic abundance goes like $\Sigma_c \propto 1/m_\chi$. On the
  other hand, for DM annihilation, the EFT validity condition reads
  $\Lambda \gtrsim 2m_\chi$. Since $\Sigma_c$ is a combination of
  couplings, we can generically write it as $\Sigma_c \sim
  1/\Lambda^2$. Therefore, by combining the validity condition with
  the relic abundance requirement, a \emph{maximum} DM mass is
  obtained for which the EFT ceases to be valid about
  $\mathcal{O}(2-5$~TeV) . Notice however that we do not expect the data
  to strongly constrain thermal DM in this regime.

\section{Conclusions}
\label{sec:concl}

In this work we have explored how strongly the combination of present data from very different experimental probes can constrain the different couplings between Dark Matter (DM) and the Standard Model (SM) particle content. The focus of the work was a bottom-up approach to understand what data itself is able to tell us regarding DM interactions, minimizing when possible any theoretical input or bias. To this end, we make use of an Effective Field Theory (EFT) approach to attain the desired model-independence at the expense of assuming that DM--SM interactions are mediated by heavy particles that can be integrated out of the theory. Furthermore, we have allowed independent couplings of DM to all SM fields so as to minimize any theoretical bias, but we have limited these interactions to flavour-conserving ones in the SM sector, since generally stronger constraints apply to them. Finally, if all possible Lorentz structures were simultaneously allowed, the parameter space would become too large to explore and constrain with present data. Thus, we restrict our analysis to Dirac DM fermions which interact with the SM constituents through independent operators of the form
\begin{equation}
c_{i,P} \left(\bar{\chi} \gamma^\mu \chi \right) \left(\bar{f_i} \gamma_\mu P f_i \right),
\end{equation}
with independent coefficients $c_{i,P}$ for all SM fermions $f_i$ and chiralities $P \equiv P_L, P_R$. 

This working model, which is not intended to reproduce any particular ultraviolet completion, provides the necessary parametrization to assess through a joint analysis the constraints on the individual interaction strengths from present data. We constrain 15 independent coefficients through a combination of 8 experimental probes comprising relic abundance, direct and indirect DM searches as well as collider limits. In addition to this general setup, we also explore how these constraints are affected when correlations between the different coefficients are allowed, or when some of the operators are forbidden. In particular, we studied the completely leptophilic, leptophobic and flavour-blind cases, as well a family-oriented scenario in which all members of the same generation share the same coupling to the DM particle.

From our global fits we find that for DM masses $m_\chi <
  20$~GeV the tension with respect to the best fit between the couplings necessary to reproduce the
  observed DM relic abundance and the upper bounds from null results
  exceeds the $3 \sigma$ level, for the general and leptophilic
  scenarios considered, while the same happens for $m_\chi \lesssim 40$~GeV
  for the leptophobic, flavour blind and family oriented cases. For
  these three cases, the $\chi^2$ rises very steeply as the DM mass
  decreases such that for $m_\chi < \mathcal{O}(10-20)$~GeV this
  tension reaches the $5 \sigma$ level, within our EFT framework. 

Furthermore we find that, due to the slightly weaker present
constraints, couplings to the second-generation lepton doublet are
preferred, whenever they are available within the model under
study. If this coupling is not available or is instead further
constrained by an assumed correlation, as is the case of the
flavour-blind scenario, a more sizable coupling to the right-handed,
up-type quark singlet is instead preferred for similar reasons. Thus,
it would be interesting to improve our present constraints on these
couplings. Finally, we also find that, since all couplings are assumed
to be independent and free to vary in the fit, the very stringent
constraints stemming from direct search experiments such as LUX imply
that either the first generation quark couplings to DM are extremely
small or that they are related to each other such that the different
contributions to these processes cancel against each other, i.e.,
$2\cQ1 + \cu + \cd \simeq 0$.

DM searches worldwide are now probing and constraining essentially all possible interaction channels between DM and the known matter constituents through extremely different and complementary and search techniques. In this context, and given our lack of a unique theoretical DM paradigm, it is important to test different DM models against present data with bottom-up approaches where theoretical biases are not imposed. Unfortunately, true and complete model independence cannot be achieved. Indeed, while EFT offers an ideal frame for this sort of studies, its adoption already enforces some assumptions, such as the decoupling nature of the mediating particle or the uniqueness of the DM candidate. Moreover, true model independence would imply the inclusion of hundreds of independent operators with different flavour and chiral structures, rendering the analysis too general to actually provide any useful information. In this work we have reduced the number of theoretical assumptions taking a first step which implies some unavoidable restrictions to the number and types of effective operators included. It would be very interesting to supplement our results with additional complementary analyses, including different Lorentz structures, different DM fields (Majorana fermions, or scalars), or with extensions beyond the EFT approach by allowing one (or several) generic light mediator(s).

\begin{acknowledgments}
We are happy to acknowledge stimulating discussions with Felix Kahlhoefer, Olga Mena, Miguel Peiro,
Pantelis Tziveloglou and Aaron Vincent.  The work of MB was
supported by the G\"oran Gustafsson Foundation. PC, EFM and PM
acknowledge financial support by the European Union through the ITN
INVISIBLES (PITN-GA-2011-289442). EFM also acknowledges support from
the EU through the FP7 Marie Curie Actions CIG NeuProbes
(PCIG11-GA-2012-321582) and the Spanish MINECO through the ``Ram\'on y
Cajal'' programme (RYC2011-07710) and the project FPA2009-09017. EFM
and PM were also supported by the Spanish MINECO through the Centro de
excelencia Severo Ochoa Program under grant SEV-2012-0249. The work of
BZ is supported by the IISN and by the Belgian Federal Science Policy
through the Interuniversity Attraction Pole P7/37. Fermilab is
operated by the Fermi Research Alliance under contract
no. \protect{DE-AC02-07CH11359} with the U.S. Department of
Energy. The work of PC was partially supported by the U.S. Department
of Energy under contract \protect{DE-SC001363}. PC and PM would like
to thank the Mainz Institute for Theoretical Physics for its
hospitality and partial support during the completion of this
work. EFM and PM thank the Aspen Center for Physics for its
hospitality and the support of the National Science Foundation grant
PHY-1066293 and the Simons Foundation for their stay there.

\end{acknowledgments}

\appendix
\section{Fitting procedure for the AMS02 positron flux data}
\label{sec:app}

As already explained in Sec.~\ref{sec:constraints}, the measurements from AMS on the positron fraction are also used to
derive upper bounds on DM annihilation cross sections. The main issue that has to be dealt with when doing so, however, are the large uncertainties on the positron and electron fluxes from astrophysical sources. It is common to use a linear combination of two power laws to parametrize them (see, e.g., Refs~\cite{Ibarra:2013zia,Bergstrom:2013jra,Kopp:2013eka}), where typically the positron flux includes an exponential cut-off at high energies. Our parametrization for the electron and positron fluxes is shown in Eq.~\eqref{eq:posiflux}. For the propagation of charged cosmic rays in the astrophysical media, we use the MED parameters defined in Ref.~\cite{Donato:2003xg}.

In principle, both DM and the astrophysical contribution to the electron and positron fluxes are affected by the solar modulation. This can be explicitly taken into account by computing the flux at the top of the atmosphere ($\oplus$) under the force field approximation, see Eq.~\eqref{eq:solarmod}. Therefore, under these assumptions our background already depends on 9 parameters
\begin{equation}
\left.
\begin{array}{lcl}
\{ C_e , \gamma_e , C_s , \gamma_s , E_s , \phi^{+} \} & \quad & \textrm{for\, positrons, \, and}  \\
\{ C_1 , \gamma_1 , C_2 , \gamma_2 , \phi^- \} & \quad & \textrm{for\, electrons. } 
\end{array}
\right.
\label{eq:posiparams}
\end{equation}
To get the total positron fraction, one would add to the background the additional contribution from DM annihilation, which in principle depends on the solar modulation parameter $\phi^+$ as well, see Eq.~\eqref{eq:posifraction}.

In principle, the best thing would be to perform a fit to the AMS02 data where, for each value of the DM mass, the minimum of the $\chi^2$ is searched for after marginalizing over the 9 parameters listed above. However, this is computationally rather expensive. Furthermore, we found that severe degeneracies take place among the different parameters, which complicates the problem even further. 

However, the problem can be considerably simplified by considering the following. In principle, the positron flux will be the most sensitive to the DM annihilation signal, while the electron flux will be mainly dominated by astrophysical backgrounds instead. This allows to reduce the number of parameters in the fit significantly: since the electron flux will be independent from the signal, it can be fitted independently and leave the parameters fixed during the fit. This is done using the parametrization in Eq.~\eqref{eq:posiflux} for the electron flux, and the publicly available data\footnote{The AMS02 data in Ref.~\cite{Aguilar:2013qda} contains only the positron fraction and fluxes, but not the electron data. Therefore, we use the Fermi LAT data in this case, which is publicly available~\cite{Ackermann:2010ij}. } on the electron flux from the Fermi LAT collaboration~\cite{Ackermann:2010ij}.
 
\begin{figure}[ht]
\centering
\includegraphics[width=0.45\textwidth,angle=0]{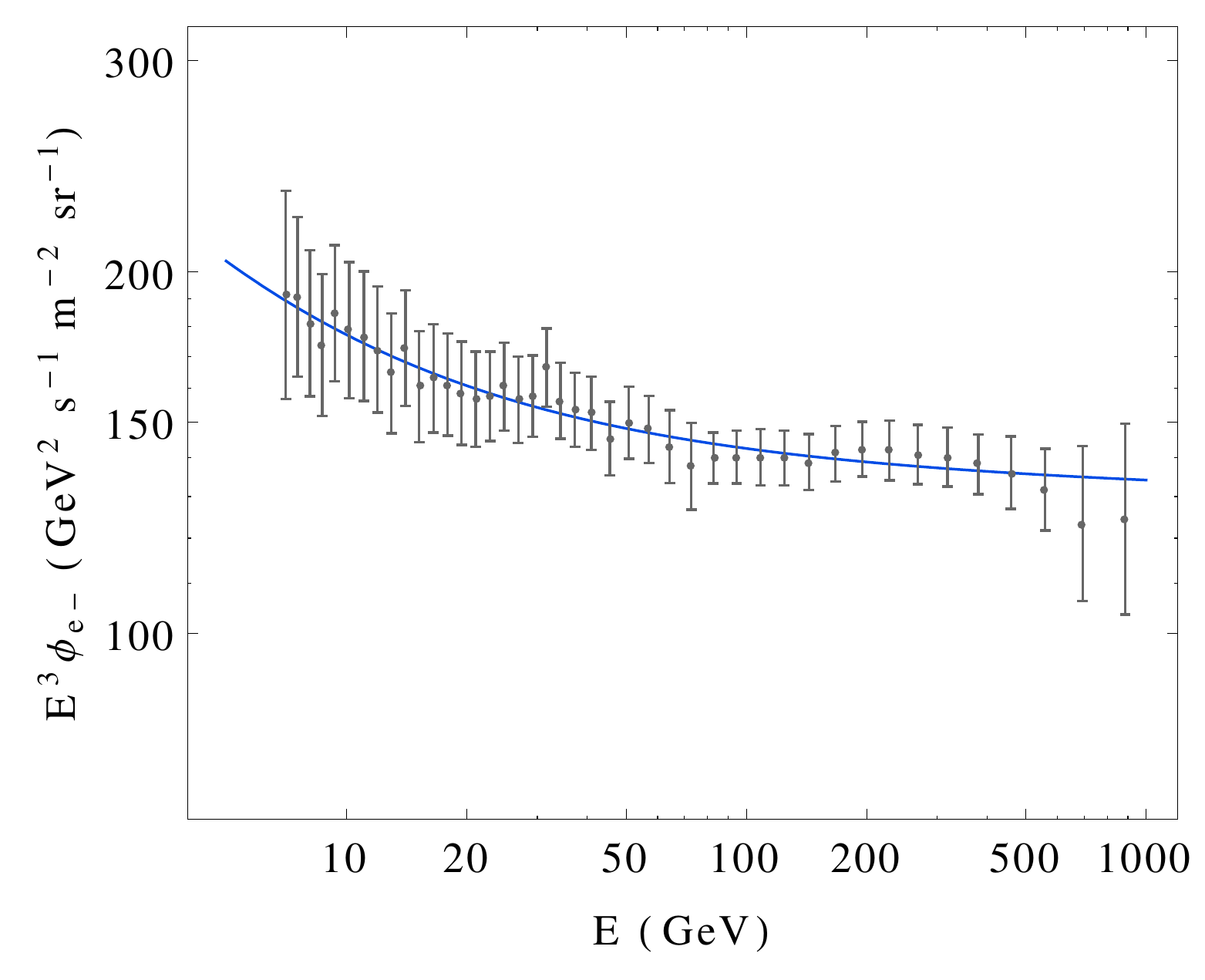}
\includegraphics[width=0.45\textwidth,angle=0]{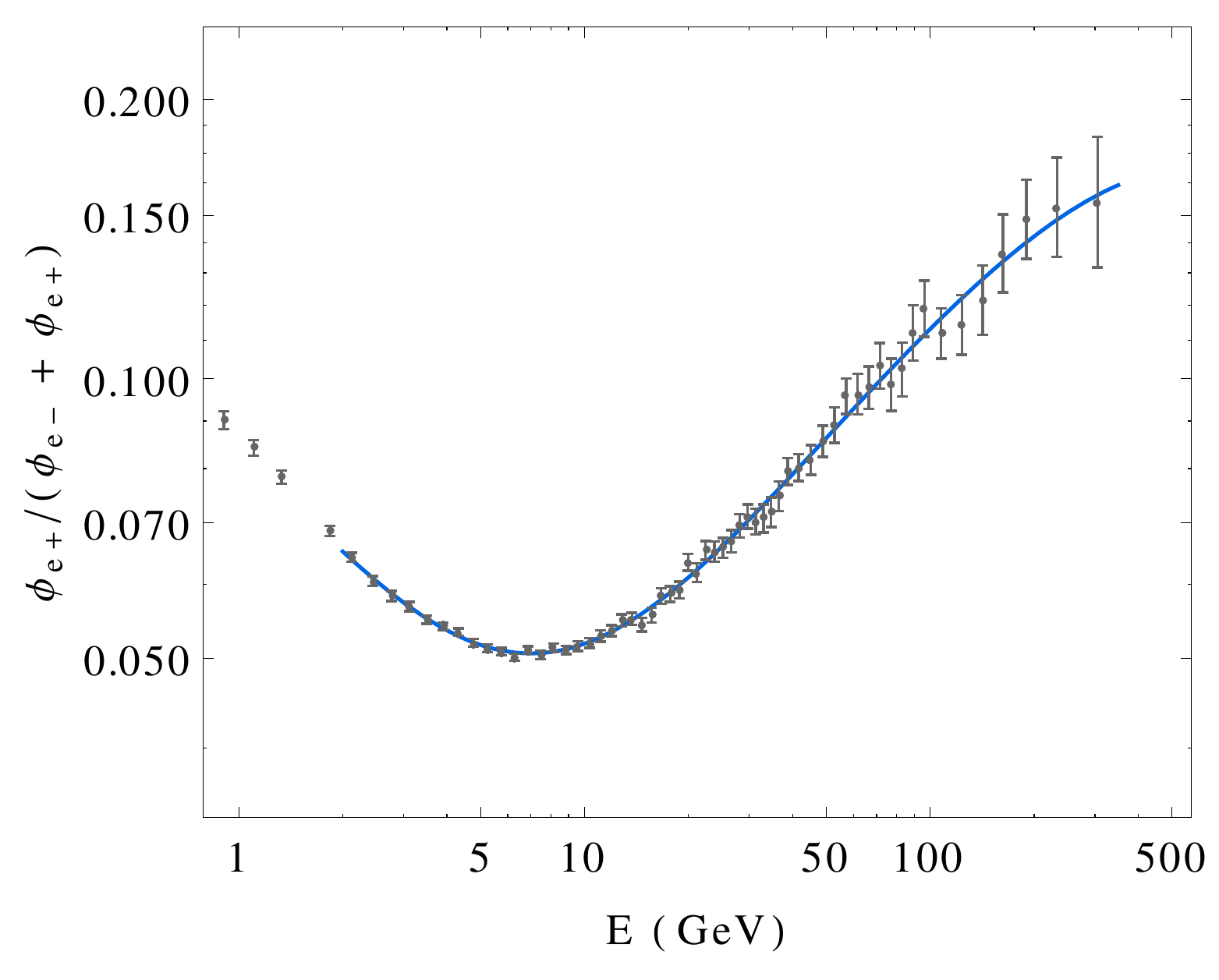}
\caption{\footnotesize{(Left) Our best-fit for the electron flux compared to Fermi LAT data. (Right) Our best-fit to the positron fraction in the absence of any contribution from DM annihilation, compared to the AMS data. See text for details. }}
\label{fig:fermiAMS}
\end{figure}

When fitting the Fermi LAT data, a $\chi^ 2$ fit is performed considering both the low-energy and high-energy data sets, between 7~GeV and 1~TeV. The resulting curve is shown in the left panel in Fig.~\ref{fig:fermiAMS}, together with the data points as extracted from Ref.~\cite{Ackermann:2010ij}. The uncertainties in each bin are computed by adding in quadrature their statistical and systematic errors.\footnote{In the case of asymmetric systematic errors, we (conservatively) take the largest value.} We find that the parameters which give a best-fit to the Fermi LAT electron flux data are
\begin{equation}
\left.
\begin{array}{rcl}
C_{1(2)} & = & 213.7 \,(140.1)~{\rm s}^{-1}\, {\rm sr}^{-1}\, {\rm m}^{-2}\, {\rm GeV}^{-1}  \\
 \gamma_{1(2)} & = & 3.7\,(3.0) \, ; \, \phi^-  =  0.0 . 
\end{array}
\right.
\end{equation}

For these values of the parameters, we find a minimum $\chi^2/\textrm{d.o.f.}= 4.7/38$.

Finally, when deriving the constraint from AMS data, the total positron flux is obtained as the addition of the astrophysical background (which depends on $C_e,C_s,\gamma_e,\gamma_s, E_s$ and $\phi^+$) and the contribution from DM annihilation. In the absence of an extra contribution from DM annihilation, we find that the following parameters give a best-fit to the positron fraction data from AMS
\begin{equation}
\left.
\begin{array}{rcl}
C_{e(s)} & = & 30.4 \,(2.0)~{\rm s}^{-1}\, {\rm sr}^{-1}\, {\rm m}^{-2}\, {\rm GeV}^{-1} ; \\
 \gamma_{e(s)} & = & 3.9 \,(2.5) \, ; \, E_s = 1086.8 {\rm GeV} ; \, \phi^+ = 0.0 ,  
\end{array}
\right.
\end{equation}
with $\chi^2/\textrm{d.o.f.} = 26.2/53 $. The positron fraction obtained with these parameters can be seen in the right panel in Fig.~\ref{fig:fermiAMS} together with the AMS data. Again in this case, the errors are taken as the sum in quadrature of the statistical and systematic errors in each bin. 

In our simulations, however, we include the DM annihilation to the positron flux and we let the parameters in Eq.~\eqref{eq:posiparams} vary during the fit. When doing so, we find that the solar modulation parameter does not have a major impact in the fit while it generates some numerical degeneracies with other parameters, which are difficult to deal with. Therefore, we also fix this parameter to the value which gives a best-fit to the AMS02 data using the background contribution alone. The rest of the parameters in Eq.~\eqref{eq:posiparams} are left free during the fit, and will be fitted for each value of the DM mass and the couplings in an independent way.

\begin{figure}[ht]
\centering
\includegraphics[width=0.45\textwidth,angle=0]{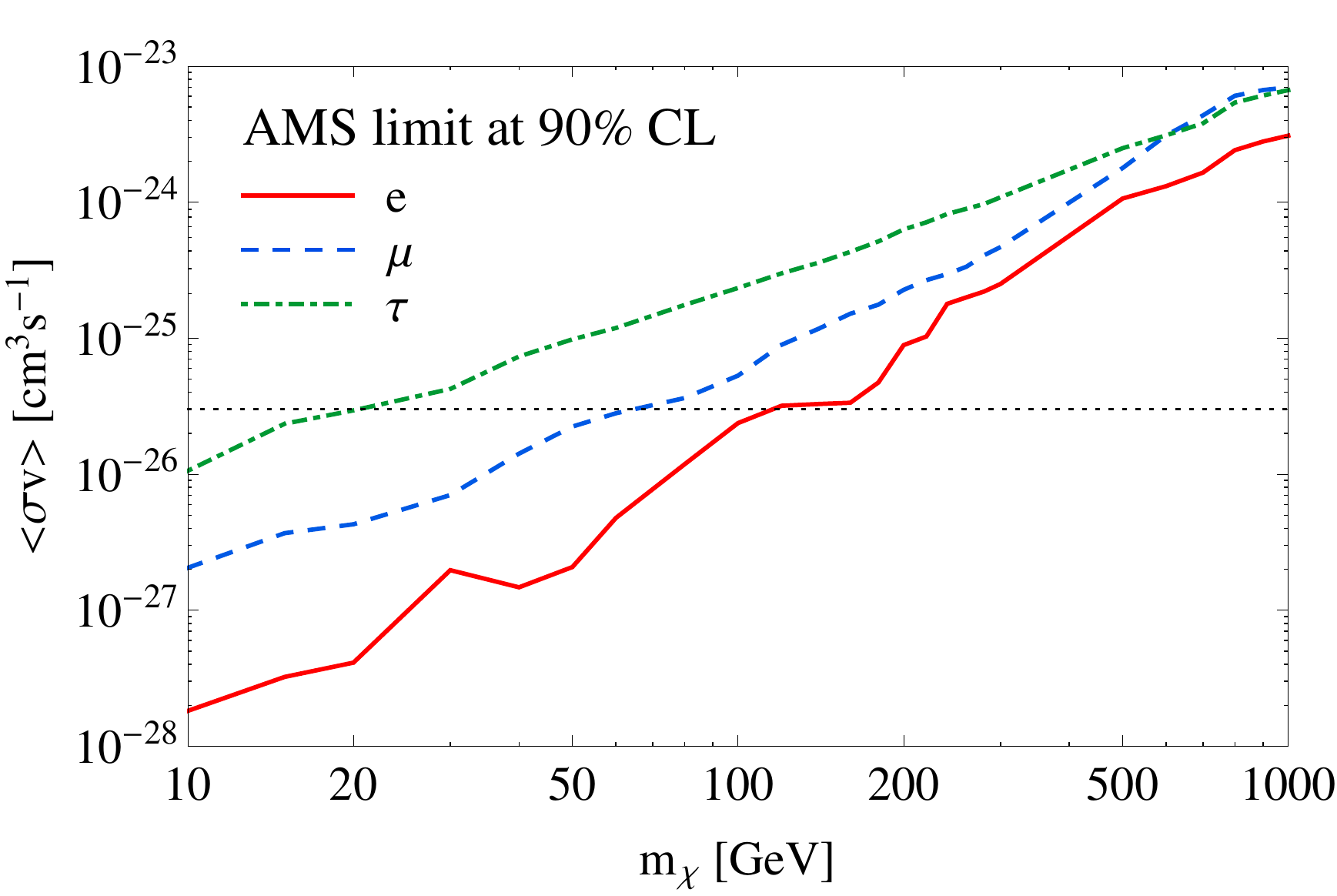}
\caption{\footnotesize{Limits to the DM interaction cross section as a function of the DM mass for several primary annihilation channels to leptons, as indicated in the legend, when only one coupling is allowed at a time. The regions above each line are excluded at 90\% CL (1 dof). The horizontal line indicates the cross section needed to satisfy relic abundance constraints. }
\label{fig:AMSlimitsL}}
\end{figure} 

\begin{figure}[ht]
\centering
\includegraphics[width=0.45\textwidth,angle=0]{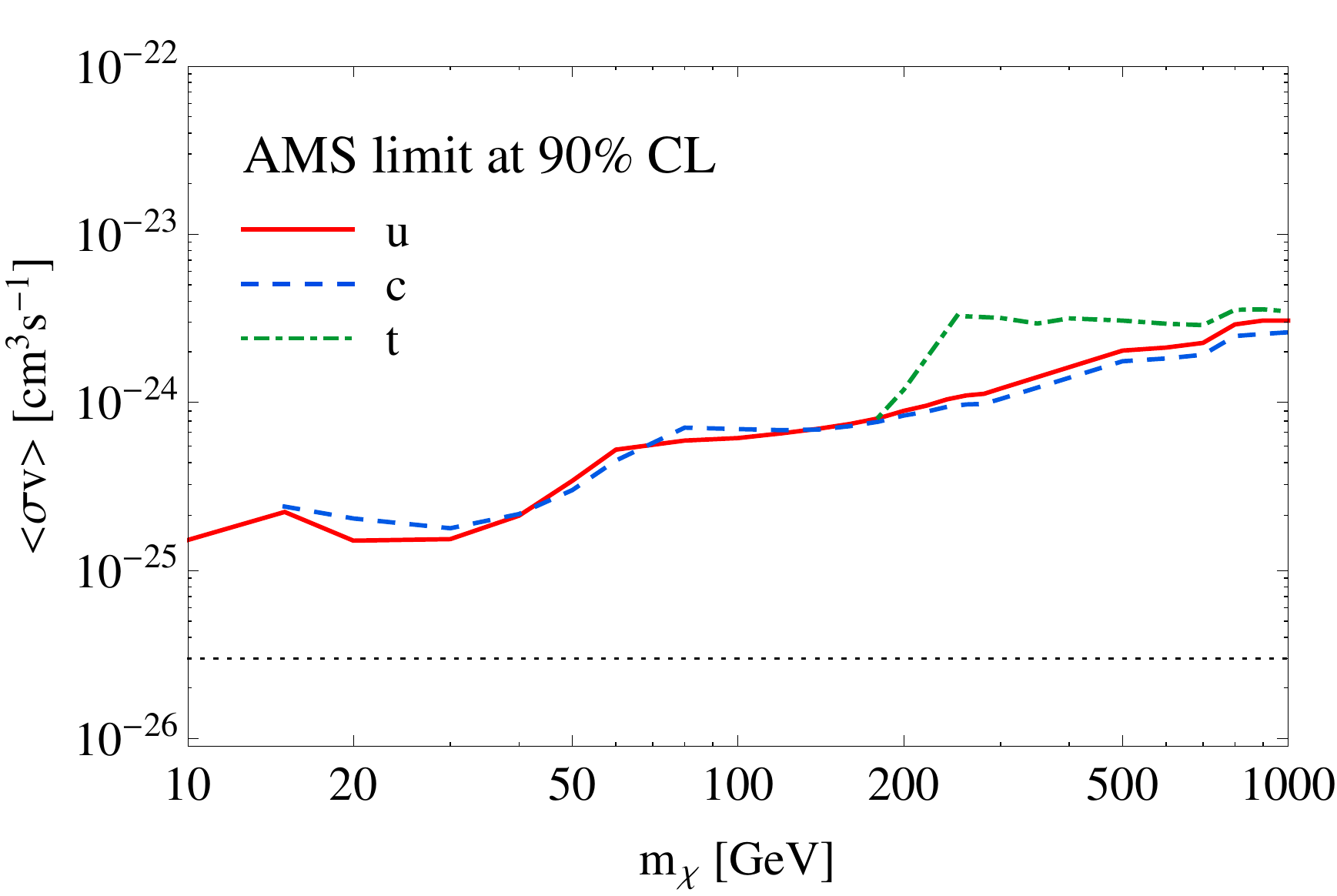}
\includegraphics[width=0.45\textwidth,angle=0]{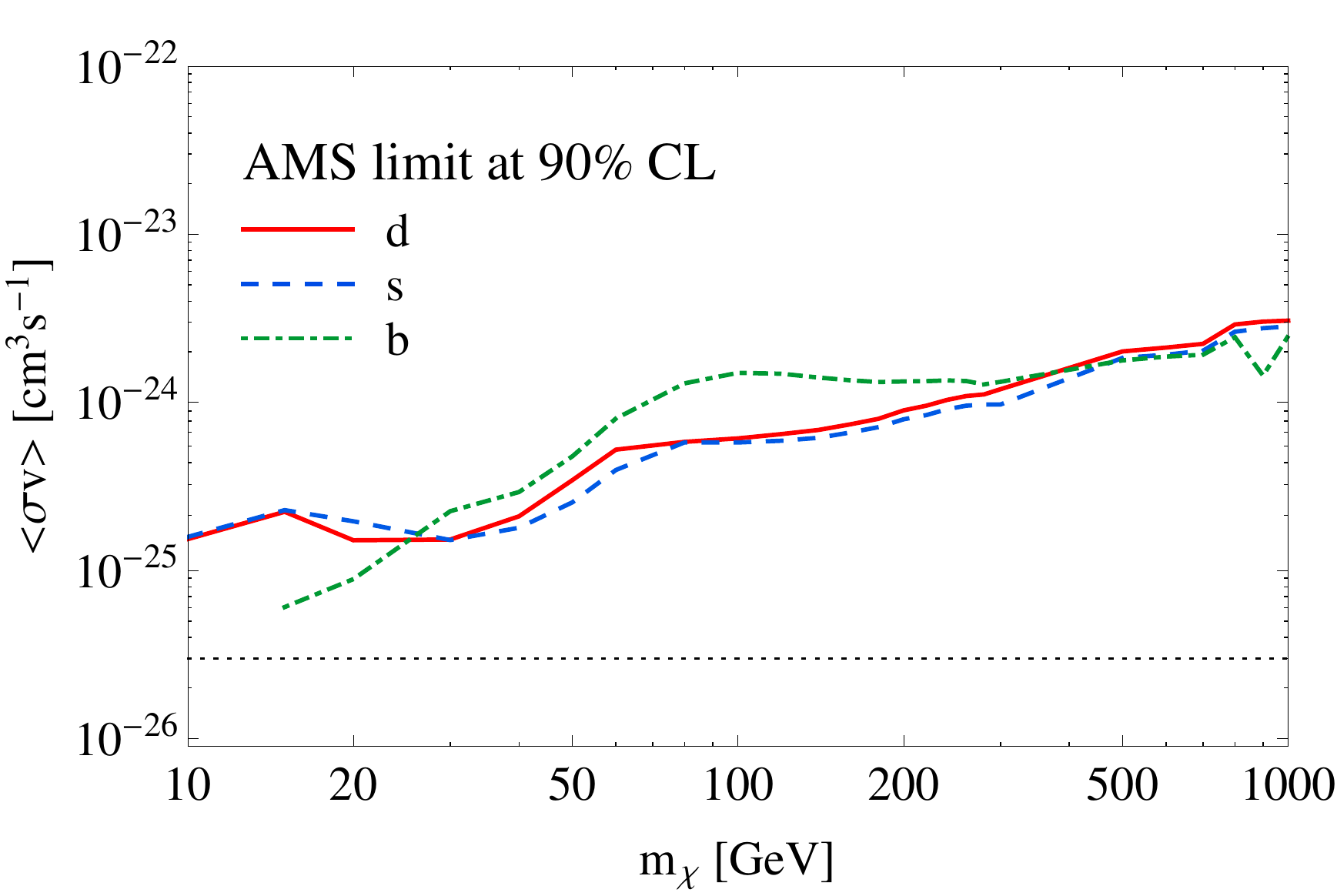}
\caption{\footnotesize{Same as Fig.~\ref{fig:AMSlimitsL} but for primary DM annihilation into quarks, as indicated in the legend.} 
\label{fig:AMSlimitsQ}}
\end{figure} 

Finally, even though in our simulations we consider several couplings between the DM and SM particles at once, it is useful to look at the limits obtained when only one coupling to a SM fermion is allowed at a time. This allows to illustrate the interplay between different data sets and where the tension in the fit for low values of the DM mass may come from. The limits on the DM-SM interaction cross section are shown in Fig.~\ref{fig:AMSlimitsL} for leptons and in Fig.~\ref{fig:AMSlimitsQ} for quarks, as a function of the DM mass. Since in our fit we are combining AMS with relic abundance constraints, a significant tension will only take place when the limit on the interaction cross section gets below the value needed for relic abundance. Therefore, one can already see from this figure that the AMS data will be most effective in constraining the couplings of the DM to electrons (for $m_\chi \lesssim 10$~GeV) and muons (for $m_\chi \lesssim 60$~GeV), but will be less efficient for other DM fermions (for instance, for taus it will only significantly affect the fit for $m_\chi\lesssim 20$~GeV).


\end{document}